\documentclass[reprint, nofootinbib, twocolumn, twoside]{revtex4-2}

\usepackage{graphicx}
\usepackage{natbib}
\usepackage{amsmath}
\usepackage{amsfonts}
\usepackage{amssymb}
\usepackage{etoolbox}

\makeatletter
\def\p@paragraph{\thesection\,\thesubsection\,}
\makeatother

\newcommand{\icmp}    {\ensuremath{\mathrm{i}}}

\newcommand{\tg}      {\ensuremath{\mathrm{tg}\,}}

\newcommand{\sgn}     {\ensuremath{\mathrm{sgn}}}
\newcommand{\intdbl}  {\ensuremath{\int}}
\newcommand{\inttrp}  {\ensuremath{\int}}
\newcommand{\differ}  {\ensuremath{\mathrm{d}}}

\newcommand{\entj}    {\ensuremath{j}}
\newcommand{\entn}    {\ensuremath{n}}

\newcommand{\jacobien}{\ensuremath{\mathrm{J}}}
\newcommand{\pscal}   {\ensuremath{\boldsymbol{.}}}
\newcommand{\ptens}   {\ensuremath{\,\otimes\,}}

\newcommand{\jfour}   {\ensuremath{\mathcal{F}}}

\newcommand{\epso}    {\ensuremath{\varepsilon_{0}}}

\newcommand{\cx}    {\ensuremath{x}}
\newcommand{\cy}    {\ensuremath{y}}
\newcommand{\cz}    {\ensuremath{z}}

\newcommand{\cX}    {\ensuremath{X}}
\newcommand{\cY}    {\ensuremath{Y}}
\newcommand{\cZ}    {\ensuremath{Z}}
\newcommand{\Ox}    {\ensuremath{Ox}}
\newcommand{\Oy}    {\ensuremath{Oy}}
\newcommand{\Oz}    {\ensuremath{Oz}}
\newcommand{\rvec}  {\ensuremath{\mathbf{r}}}
\newcommand{\rmod}  {\ensuremath{r}}

\newcommand{\temps} {\ensuremath{t}}
\newcommand{\tpre}  {\ensuremath{\temps_{0}}}
\newcommand{\tmes}  {\ensuremath{\temps_{1}}}
\newcommand{\tecr}  {\ensuremath{\temps_{2}}}

\newcommand{\inist} {\ensuremath{0}}
\newcommand{\masse} {\ensuremath{m}}

\newcommand{\freq}  {\ensuremath{{\omega}}}
\newcommand{\freqo} {\ensuremath{\freq_{0}}}
\newcommand{\vlum}    {\ensuremath{c}}

\newcommand{\mvit}    {\ensuremath{v}}

\newcommand{\ouvas}    {\ensuremath{\mathrm{A}}}
\newcommand{\surface}  {\mbox{S}}
\newcommand{\surfouva} {\ensuremath{\surface(\ouvas)}}
\newcommand{\ouva}     {\ensuremath{\mathcal{A}}}

\newcommand{\ouvcin}   {\ensuremath{\mathcal{K}_{0}}}

\newcommand{\apersize}{\ensuremath{\Delta}}

\newcommand{\fente}   {\ensuremath{\mathrm{R}}}

\newcommand{\larg}    {\ensuremath{a}}
\newcommand{\haut}    {\ensuremath{b}}

\newcommand{\distsodi} {\ensuremath{d_{0}}}
\newcommand{\distdipo} {\ensuremath{d}}
\newcommand{\vdistdipo}{\ensuremath{\mathbf{ d}}}

\newcommand{\lgond}    {\ensuremath{\lambda}}

\newcommand{\vkond}    {\ensuremath{\mathbf{k}}}
\newcommand{\vkondini} {\ensuremath{\vkond_{\inist}}}

\newcommand{\mkond}  {\ensuremath{k}}
\newcommand{\mkondini} {\ensuremath{\mkond_{\inist}}}
\newcommand{\kox}      {\ensuremath{\mkond_{0\cx}}}
\newcommand{\koy}      {\ensuremath{\mkond_{0\cy}}}
\newcommand{\koz}      {\ensuremath{\mkond_{0\cz}}}
\newcommand{\kx}     {\ensuremath{{\mkond_{\cx}}}}
\newcommand{\ky}     {\ensuremath{{\mkond_{\cy}}}}
\newcommand{\kz}     {\ensuremath{{\mkond_{\cz}}}}
\newcommand{\amplik} {\ensuremath{\mathcal{U\,}_{\mathrm{WO}}}}

\newcommand{\vargc}  {\ensuremath{C}}
\newcommand{\facincl}{\ensuremath{\Omega}}
\newcommand{\debket}    {\ensuremath{\left|\,}}
\newcommand{\finket}    {\ensuremath{\;\right>\,}}
\newcommand{\debbra}    {\ensuremath{\,\left<\,}}
\newcommand{\finbra}    {\ensuremath{\,\right|}}
\newcommand{\debpscal}  {\ensuremath{\,\left<\,}}
\newcommand{\midpscal}  {\ensuremath{\left|\,}}
\newcommand{\finpscal}  {\ensuremath{\right.\right>\,}}

\newcommand{\debvalop}  {\ensuremath{\,\left<\,}}
\newcommand{\midvalopa} {\ensuremath{\,\left|\left.}}
\newcommand{\midvalopb} {\ensuremath{\,\right|\,}}
\newcommand{\finvalop}  {\ensuremath{\right.\right>\,}}

\newcommand{\probcon}  {\ensuremath{P}}

\newcommand{\ilkoeoq} {\ensuremath{^{}_{\cZ[\vkondini]\!\!\!}}}

\newcommand{\ilkoeoqe}{\ensuremath{^{}_{\cZ[\vkondini],\Eularop_{0}\!\!\!}}}
\newcommand{\irkoeoqe}{\ensuremath{^{}_{\!\!\cZ[\vkondini],\Eularop_{0}}}}

\newcommand{\ilkoeoz} {\ensuremath{\,^{}_{\vkondini\!\!\!}}}
\newcommand{\irkoeoz} {\ensuremath{^{}_{\!\!\vkondini}}} 

\newcommand{\ilkeoq}  {\ensuremath{^{}_{\;\cZ[\vkond]}\!\!}}
\newcommand{\irkeoq}  {\ensuremath{^{}_{\!\!\cZ[\vkond]}}}
\newcommand{\ilkeoqe} {\ensuremath{^{}_{\;\cZ[\vkond],\Eularop}\!\!}}
\newcommand{\irkeoqe} {\ensuremath{^{}_{\!\!\cZ[\vkond],\Eularop}}}

\newcommand{\irkeoqz} {\ensuremath{^{}_{\!\!\cz[\vkond]}}}

\newcommand{\ilkoeoqz}{\ensuremath{^{}_{\;\;\;\,\cz[\vkondini]}\!}}
\newcommand{\irkoeoqz}{\ensuremath{^{}_{\!\!\cz[\vkondini]}}}

\newcommand{\ilkeoz}  {\ensuremath{\,^{}_{\vkond\!\!\!}}}
\newcommand{\ilkeozp} {\ensuremath{\,^{}_{\vkond'\!\!\!}}}
\newcommand{\irkeoz}  {\ensuremath{^{}_{\!\!\vkond}}}
\newcommand{\irkeozp} {\ensuremath{^{}_{\!\!\vkond'}}}

\newcommand{\irkoeiz} {\ensuremath{^{}_{\!\!\vkondini,\azaxeli}}}
\newcommand{\ilkeuz}  {\ensuremath{\,^{}_{\!\!\vkond,\Eularop\!\!\!}}}
\newcommand{\irkeuz}  {\ensuremath{^{}_{\!\!\vkond,\Eularop}}}

\newcommand{\veclinp}   {\ensuremath{\mathbf{e}}}

\newcommand{\xpol}      {\ensuremath{x}}
\newcommand{\ypol}      {\ensuremath{y}}

\newcommand{\spinz}     {\ensuremath{\sigma}}
\newcommand{\helic}     {\ensuremath{\xi}}
\newcommand{\helicin}   {\ensuremath{\helic_{0}}}
\newcommand{\ellipin}   {\ensuremath{\eta_{0}}}
\newcommand{\ilinp}     {\ensuremath{l}}
\newcommand{\ilinq}     {\ensuremath{L}}

\newcommand{\etat}    {\ensuremath{\psi}}
\newcommand{\etatm}   {\ensuremath{\varphi}}
\newcommand{\etcospin}{\ensuremath{\chi}}

\newcommand{\modspin}   {\ensuremath{{s}}}
\newcommand{\spinun}    {\ensuremath{{1}}}

\newcommand{\kisin}  {\ensuremath{\etcospin^{(\modspin)}_{\textup{in}}}}
\newcommand{\kiuin}  {\ensuremath{\etcospin^{(\spinun)}_{\textup{in}}}}
\newcommand{\kisoutk}{\ensuremath{\etcospin^{(\modspin)}_{\textup{out}}(\vkond)}}
\newcommand{\kiuoutk}{\ensuremath{\etcospin^{(\spinun)}_{\textup{out}}(\vkond)}}

\newcommand{\etaelli}  {\ensuremath{\widetilde{\chi}}}

\newcommand{\etatiz}  {\ensuremath{\mkond_{\inist\cz}}}

\newcommand{\opcrea}   {\ensuremath{\hat{a}^{\dagger}}}

\newcommand{\chelec}   {\ensuremath{\hat{\mathbf{E}}}}
\newcommand{\opchelcre}{\ensuremath{\chelec^{(-)}}}
\newcommand{\opchelani}{\ensuremath{\chelec^{(+)}}}
\newcommand{\opchelpm} {\ensuremath{\chelec^{(\pm)}}}
\newcommand{\chpotv}   {\ensuremath{\hat{\mathbf{A}}}}
\newcommand{\oppotvcre}{\ensuremath{\chpotv^{(-)}}}
\newcommand{\oppotvani}{\ensuremath{\chpotv^{(+)}}}
\newcommand{\oppotvpm} {\ensuremath{\chpotv^{(\pm)}}}
\newcommand{\ketvide}  {\ensuremath{\debket\!\textup{vac}\!\finket}}
\newcommand{\bravide}  {\ensuremath{\debbra\!\textup{vac}\!\finbra}}

\newcommand{\sig}     {\ensuremath{\sigma}}
\newcommand{\sigz}    {\ensuremath{\sig_{\cz}}}

\newcommand{\vamkond} {\ensuremath{K}}
\newcommand{\vavkond} {\ensuremath{\mathbf{K}}}
\newcommand{\vaspinz} {\ensuremath{[\Sigma]_{\cZ[\vavkond]}}}
\newcommand{\vaspinzc} {\ensuremath{[\Sigma]_{\cZ[\vavkond]}|\vavkond=\vkond}}
\newcommand{\vaspinzh} {\ensuremath{[\Sigma]_{\vavkond}|\vavkond=\vkond}}
\newcommand{\vaxpxizh} {\ensuremath{[X]_{\vavkond,\Eularop}|\vavkond=\vkond}}

\newcommand{\vadifx}  {\ensuremath{\Theta_{\cx}}}
\newcommand{\vadify}  {\ensuremath{\Theta_{\cy}}}
\newcommand{\gaussmk} {\ensuremath{\widetilde{\delta}^{\,\Delta\mkond}}}

\newcommand{\cs}      {\ensuremath{z}}

\newcommand{\pdfp}     {\ensuremath{f}}
\newcommand{\pdfpj}   {\ensuremath{F}}
\newcommand{\intens}   {\ensuremath{I}}

\newcommand{\indtran}  {\ensuremath{\mathrm{T}}}
\newcommand{\indlong}  {\ensuremath{\mathrm{L}}}
\newcommand{\Proj}     {\ensuremath{\hat{\mathrm{P}}}}
\newcommand{\Projtran} {\ensuremath{\Proj_{\indtran}}}
\newcommand{\Projlong} {\ensuremath{\Proj_{\indlong}}}

\newcommand{\opedif}    {\ensuremath{\hat{\mathrm{D}}}}
\newcommand{\opredu}    {\ensuremath{\hat{\mathrm{F}}}}

\newcommand{\opredutran} {\ensuremath{\opredu_{\indtran}}}
\newcommand{\opredulong} {\ensuremath{\opredu_{\indlong}}}

\newcommand{\adjoint}    {\ensuremath{{}^{^{\textrm{\scriptsize$\dagger$}}}}}
\newcommand{\fred}     {\ensuremath{\widetilde{\delta}}}
\newcommand{\fredAt}   {\ensuremath{\fred_{\indtran}^{\ouvas}}}
\newcommand{\fredDzl}  {\ensuremath{\fred_{\indlong}^{\Delta\cz}}}

\newcommand{\fredA}    {\ensuremath{\fred^{\ouva}}}
\newcommand{\fredcin} {\ensuremath{\fred^{\,\ouvcin}}}

\newcommand{\norm}   {\ensuremath{{N}}}
\newcommand{\consn}  {\ensuremath{C}}

\newcommand{\tdiftran}{\ensuremath{T}}
\newcommand{\tdiflong}{\ensuremath{L}}
\newcommand{\angdif}  {\ensuremath{\theta}}
\newcommand{\angdev}  {\ensuremath{\theta}}
\newcommand{\fgeomang}{\ensuremath{\Gamma}}
\newcommand{\eulea}    {\ensuremath{\alpha}}
\newcommand{\euleb}    {\ensuremath{\beta}}
\newcommand{\eulec}    {\ensuremath{\gamma}}
\newcommand{\eulera}   {\ensuremath{\eulea_{1}}}
\newcommand{\eulerb}   {\ensuremath{\eulea_{2}}}
\newcommand{\eulerc}   {\ensuremath{\eulea_{3}}}

\newcommand{\eulapre}  {\ensuremath{\phi}}
\newcommand{\eulanuy}  {\ensuremath{\theta}}

\newcommand{\azaxel}   {\ensuremath{\zeta}}
\newcommand{\azaxeli}  {\ensuremath{\azaxel_{0}}}
\newcommand{\Eulapre}  {\ensuremath{\Phi}}
\newcommand{\Eulanuy}  {\ensuremath{\Theta}}
\newcommand{\Eularop}  {\ensuremath{\Psi}}

\newcommand{\oprotos}  {\ensuremath{\mathcal{R}}}
\newcommand{\oprotsp}  {\ensuremath{\hat{\mathrm{R}}}}
\newcommand{\mrotd}    {\ensuremath{d}}

\begin{document}

\title{
 Position measurement and the Huygens-Fresnel principle:
 a quantum model of Fraunhofer diffraction for polarized pure states
%
%
%
%
%
%
}

\author{Bernard Fabbro}
\email[ ]{bernard.fabbro@cea.fr}
\affiliation
 {IRFU,CEA, Universit\'e Paris-Saclay, F-91191 Gif-sur-Yvette, France\\}


\begin{abstract}
 In most theories of diffraction by a diaphragm, the amplitude of the diffracted
 wave, and hence the position wave function of the associated particle, is
 calculated directly without prior calculation of the quantum state.
 Few models express the state of the particle to then deduce the position and
 momentum wave functions related to the diffracted wave.
 We present a model of this type for Fraunhofer diffraction.
 The diaphragm is assumed to be a device for measuring
 the three spatial coordinates of the particles passing through the aperture.
 A matrix similar to the S-matrix of the scattering theory describes the process
 which turns out to be more complex than a simple position measurement.
 Some predictions can be tested.
 The wavelets emission involved in the Huygens-Fresnel principle
 occurs from several neighboring wavefronts instead of just one,
 causing typical damping of the diffracted wave intensity.
 An angular factor plausibly accounts for
 the decrease in intensity at large diffraction angles,
 unlike the obliquity factors of the wave optics theories.
 The position measurement modifies the polarization states and for an incident
 photon in an elliptically polarized pure state, the ellipse axes
 can undergo a rotation which depends on the diffraction angles.
\end{abstract}

\maketitle


\section{Introduction}
{
 Quantum mechanics is involved in many studies on diffraction.
 Since the first quantum theory of Fraunhofer diffraction
 by a grating \cite{EpsEhr}, several models have emerged,
 using the formalism of path integrals \cite{PathInt,BaBas,Beau},
 the calculation of trajectories in the framework of
 hidden variables theories \cite{PhDH,SBMA}
 or the resolution of the wave equation combined
 with the use of the Kirchhoff integral \cite{Wub}.
 In more recent studies, various topics are discussed such as the effects
 of diffraction on the transmission of information
 in quantum optical systems \cite{Lupo},
 the role of the quantum behavior of the diaphragm electrons
 in diffraction of light by a small hole \cite{JuKelb},
 the interactions between the quantum states of different modes
 in diffracted Gaussian beams \cite{XiLan},
 the connection between orbital angular momentum transfer and helicity
 in the diffraction of light \cite{Deepa}.
 
 However, one question does not seem to have received much attention:
 the possibility of starting from the postulates of quantum mechanics to treat
 diffraction by a diaphragm as a consequence of a measurement of the position
 of the particle associated with the wave as it passes through the aperture.
 The first model based on this approach relates to the measurement
 of one transverse coordinate and provides the same predictions as those
 of wave optics for the case of Fraunhofer diffraction with slits \cite{Marcella}.
 Afterwards, several aspects of this model were discussed \cite{RotBou}.
 More recently, quantum trajectories has been used to describe the motion
 of the particle after the measurement of one transverse coordinate in a model
 giving predictions for Fraunhofer and Fresnel diffractions by a slit \cite{JoMat}.
 There does not seem to have been any other publications on this issue so far.

 In the model presented below, we start from the observation
 that the detection of a particle in the far field region beyond a diaphragm
 provides a measurement of its momentum.
 Then, we assume that the distribution of this momentum results
 from a measurement of the three spatial coordinates of the particle
 during its passage through the aperture and that this position measurement
 has an effect on the polarization if the particle has spin.
 The change in momentum and polarization is described
 by a "diffraction matrix" similar to the S-matrix
 of the scattering theory \cite{LLrel}.
 Although this model only applies to the far field,
 it nevertheless provides specific predictions
 about the Huygens-Fresnel principle, the diffraction at large angles
 and, in the case of light, the polarization of the photons
 detected beyond the diaphragm.

 We present the model in Sec. $\!\!\!$~\ref{sec:etcontex}.
 Next, some predictions regarding intensity and polarization measurements
 are described in Sec.  $\!\!\!$~\ref{sec:predreli}.
 Finally, we conclude in Sec. $\!\!\!$~\ref{sec:conclu}.
 
}

\section{Quantum model of Fraunhofer diffraction by an aperture}
{
\label{sec:etcontex}

\subsection{Measurement of quantities related to the detected particles.}
{
\label{sub:aspexp}

\paragraph{Experimental setup and first assumptions.}
{
 The model applies for an experimental setup
 with the following characteristics (Fig. \!\!~\ref{fig:dispexa}).
 The diaphragm is a plane assumed to be of zero thickness and perfectly opaque.
 The aperture, of finite area, can be of any shape
 and possibly formed of several parts.
 The origin of the laboratory frame of reference $(O;x,y,z)$ is located at
 the aperture and the $(\Ox,\Oy)$ plane is that of the diaphragm.
 The source is located on the $\cz$ axis and can emit non-relativistic particles
 or photons.
 Detectors placed beyond the diaphragm measure the local counting rate
 and possibly the polarization.
 The position of a detection point is denoted
 by its radius-vector $\vdistdipo$.
%
\begin{figure}[t!]\hspace{-0.75cm}
\begin{minipage}[t]{8cm}
\resizebox{1.075\textwidth}{!}{%
\includegraphics{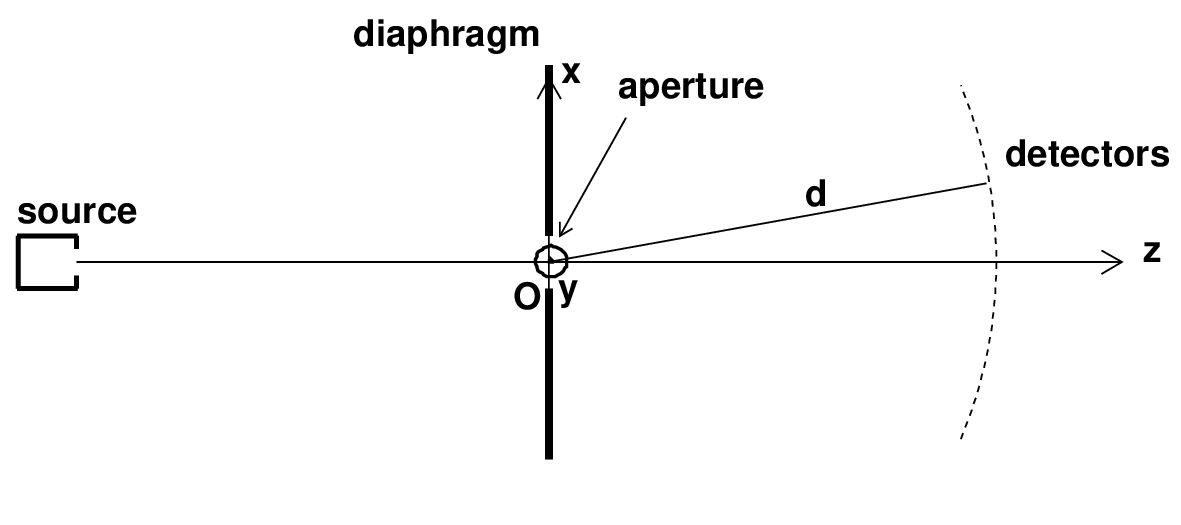}
}
\vspace{-0.75cm}
\caption{Experimental setup and laboratory frame of reference
 (right-handed coordinate system).}
\label{fig:dispexa}
\end{minipage}
\end{figure}

 We consider the case of a low intensity source and we assume that there is
 neither creation nor annihilation of particles during the passage
 of the wave through the aperture.
 We can then individually assign a quantum state to each non-relativistic
 particle and a one-photon state of the electromagnetic field to each photon,
 both for the incident wave than for the diffracted wave.
 Moreover, it can be assumed that the particle does not interact with the
 experimental device and therefore behaves like a free particle
 when it moves between the source and the diaphragm and
 between the diaphragm and the detectors.

 Finally, we consider the case where the source-diaphragm and diaphragm-detector
 distances are large enough for the aperture to be viewed as a point from the
 detectors and for the incident wave to be close to a plane wave when it arrives
 at the aperture.
 For simplicity, this plane wave is supposed to be monochromatic
 with wave vector $\vkondini$ in the direction of the $\cz$ axis.
\\

}

\paragraph{Measurement of the momentum of the detected particles.}
{
 From the above assumptions and conditions, we can assign the momentum
 $\hbar\vkondini$ to the incident particle and a momentum $\hbar\vkond$ such that:
\begin{eqnarray}\vkond \;=\;(\mkond/\distdipo)\;\vdistdipo
\label{eq:detadn} \end{eqnarray}
 to the particle detected at point of radius-vector $\vdistdipo$,
 provided that the modulus $\mkond$ is measured.
 However, no significant difference between the wavelength
 of the diffracted wave and that of the incident wave
 is observed in diffraction experiments with a diaphragm. Hence:
\begin{eqnarray}\mkond\,\simeq\,\mkondini,
\label{eq:contrax} \end{eqnarray}
 which is in accordance with kinematics because the particle transfers
 a very small part of its energy to the diaphragm.
 So it is not required to determine $\mkond$ by a special measurement.
 Furthermore, the part of the diffracted wave returning from the aperture
 to the region where the source is located is very weak.
 For simplicity, we assume that the momentum of the particle associated
 with the diffracted wave is always such that:
\begin{eqnarray} \kz\,>\,0.
\label{eq:contray} \end{eqnarray}

 The relations (\!\!~\ref{eq:detadn}), (\!\!~\ref{eq:contrax}) and
 (\!\!~\ref{eq:contray}) imply that it is possible to measure
 the momentum probability density function (p.d.f.)
 of the particle after its passage through the aperture
 on condition of being in diffraction at infinity.
 The measurement can be performed, for example,
 by arranging detectors on a hemisphere of center $O$
 and radius $\distdipo$ in the half-space $\cz>0$.
 The radius must be such that $\apersize\ll\distdipo$,
 where $\apersize$ is the size of the aperture,
 otherwise (\!\!~\ref{eq:detadn}) cannot be used.
 The Fraunhofer diffraction criterion, that is:
 $\apersize^2/(\lgond\distdipo)\,\ll\,1$ \cite{BoWo,Soma,LLangdif},
 is then satisfied if $\distdipo$ is large enough,
 whatever the value of $\lgond/\apersize$.
\\

}

\paragraph{Measurement of the polarization of the detected particles.}
{
 The polarization measuring device (analyzer for photons, Stern and Gerlach
 apparatus for atoms, etc...) is placed in front of the detector
 which is located, given (\!\!~\ref{eq:detadn}), in the direction
 of the momentum $\hbar\vkond$ of the detected particle.
 The measurement therefore gives the probabilities of the eigenvalues
 of the spin component on a quantization axis $\cZ[\vkond]$
 which must be chosen with respect to a coordinate system
 $\{\cx[\vkond],\cy[\vkond],\cz[\vkond]\}$ attached to the detected particle.
 Finally, it is possible to measure, on a particle of spin $\modspin$,
 the probability of finding the result $\spinz$ for its spin component
 on a $\cZ[\vkond]$ axis {\it if} the measurement of its momentum gives
 the result $\hbar\vkond$. It is therefore a {\it conditional} probability.
 
 By convention, the coordinate system attached to the incident particle
 is the laboratory frame of reference
 (Fig. \!\!~\ref{fig:dispexa}) whose $\cz\equiv\cz[\vkondini]$ axis
 is in the direction of the momentum $\hbar\vkondini$.
 For the detected particle, we choose the coordinate system obtained
 from the laboratory frame of reference by the rotation
 $\oprotos(\eulapre,\eulanuy,0)$ where the Euler angles are defined according to
 the z-y-z convention, so that $\eulapre$ and $\eulanuy$ are respectively
 the azimuth and the polar angle of $\vkond$. Hence:
\begin{eqnarray} \ilinp[\vkond]=\oprotos(\eulapre,\eulanuy,0)\,
 \ilinp[\vkondini]\,,\hspace{0.25cm}\ilinp=\cx,\cy,\cz;
 \hspace{0.5cm}\cz[\vkond]\parallel\vkond. \hspace{0.75cm}
\label{eq:rotkok} \end{eqnarray}
 The zero value of the third Euler angle defines a choice of the directions
 of the $\cx[\vkond]$ and $\cy[\vkond]$ axes in the transverse plane to $\vkond$
 such that the coordinate system attached to the detected particle
 in the case $\eulapre=\eulanuy=0$ is coincident
 with the laboratory frame of reference.

 Two very different cases arise concerning the quantization axis. For a particle
 of non-zero mass, this axis can be chosen in any direction. There is then
 an infinite number of possible $\cZ[\vkond]$ axes for each vector $\vkond$.
 On the other hand, for a particle of zero mass, the quantization axis
 must be in the direction of the momentum because the only spin component
 eigenstates are the helicity states \cite{LLrel}.
 There is then only one possibility which is $\cZ[\vkond]=\cz[\vkond]$,
 according to the above convention.

}

}

\subsection{Diffraction operator.}
{
\label{sub:diffrop}

\paragraph{Measurement of the position of the incident particles.}
{
 Since it is possible to measure the momentum p.d.f. and the polarisation
 of the particles associated with the diffracted wave at infinity,
 we can consider the construction of a quantum model
 whose purpose is to provide the expressions of these quantities.
 The model proposed here is based on the assumption that each incident particle
 undergoes a position measurement as it passes through the aperture.
 The detection of a particle beyond the diaphragm can indeed be considered
 as proof that it effectively passed through the aperture and was therefore
 localized at this place during a short period of time with a precision
 of the order of the size of the aperture \cite{Heisen}.
 For simplicity, we consider that the localization occurs instantaneously.
 We then assume that the source emits a particle at time $\tpre$,
 that this particle passes through the aperture at time $\tmes$
 and that it is detected at time $\tecr$.
 The time $\tmes$ can then be interpreted as the time
 of the change of state caused by a position quantum measurement
 carried out by the diaphragm
 and the purpose of the model is to build a {\it diffraction operator}
 which describes this change of state.
\\

}

\paragraph{Using S-matrix theory formalism.}
{
 The quantum state of the particle of spin $\modspin$ at time $\temps$ is
 assumed to be a pure state denoted $\debket\etat^{(\modspin)}(\temps)\finket$.
 Since the incident wave is close to a monochromatic plane wave  with wave vector
 $\vkondini$ and given (\!\!~\ref{eq:contrax}), the incident particle and the
 particle associated with the diffracted wave are
 in an energy state close to the eigenstate of eigenvalue $\hbar\freqo$,
 where $\freqo=\vlum\,(\,\hbar^{-2}\masse^2\vlum^2+{\mkondini}^2\,)^{1/2}$.
 The initial and final states are therefore described with a good approximation
 by stationary states of the form:
\begin{eqnarray}\!\!\begin{array}{rcl}\displaystyle
 \tpre<\temps<\tmes\!:\hspace{0.25cm}
 \debket\etat_{\textup{in}}^{(\modspin)}(\temps)\!\finket\!\!&\simeq&
 \exp(-\icmp\freqo\temps)\,\debket\etatm^{(\modspin)}_{\textup{in}}\finket\!,
\vspace{0.2cm} \\ \displaystyle
 \tmes<\temps<\tecr\!:\hspace{0.25cm}
 \debket\etat_{\textup{out}}^{(\modspin)}(\temps)\!\finket\!\!&\simeq&
 \exp(-\icmp\freqo\temps)\,\debket\etatm^{(\modspin)}_{\textup{out}}\finket\!,
\end{array} \hspace{0.55cm}
\label{eq:etasymp} \end{eqnarray}
 where $\debket\etatm^{(\modspin)}_{\textup{in}}\finket$ and
 $\debket\etatm^{(\modspin)}_{\textup{out}}\finket$ are time-independent states.
 Since time dependence only appears in global phase factors,
 knowing the exact values of $\tpre$, $\tmes$ and $\tecr$ is not essential
 and, as in the S-matrix theory, we consider a diffraction operator
 $\opedif^{(\modspin)}$ which projects the initial time-independent state
 on the final time-independent state (called "initial state" and "final state"
 in the following). The change of state is expressed by:
\begin{eqnarray} \debket\etatm^{(\modspin)}_{\textup{out}}\finket=
 \left[\,\norm^{(\modspin)}\,\right]^{-1/2}\; \opedif^{(\modspin)}
 \debket\etatm^{(\modspin)}_{\textup{in}}\finket,
\label{eq:fgenera} \end{eqnarray}
 where $\norm^{(\modspin)}$ is the normalization factor:
\begin{eqnarray}\norm^{(\modspin)}\equiv
 \debvalop\etatm^{(\modspin)}_{\textup{in}}
 \midvalopa\opedif^{(\modspin)}\adjoint\,
 \opedif^{(\modspin)}\midvalopb\etatm^{(\modspin)}_{\textup{in}}\finvalop.
\label{eq:fgenerb} \end{eqnarray}
 All the information on the "particle-diaphragm interaction"
 is contained in the matrix elements of the diffraction operator
 from which we can get the transition amplitudes between the initial state
 and the final momentum and spin component eigenstates.
 Since we only consider one-particle states, these eigenstates
 are represented by the state vectors:
\begin{eqnarray}\begin{array}{lrcl}
 \modspin = 0 :  \hspace{0.25cm}  & \opcrea(\vkond)\ketvide
 &\!\!=\!\!& \debket\vkond\finket,
\vspace{0.2cm}\\
 \modspin\neq 0 : \hspace{0.25cm} & 
 \opcrea\!\left(\vkond,[\spinz]_{\cZ[\vkond]}\right)\ketvide
 &\!\!=\!\!& \debket\vkond\finket\!\!\ptens\!\debket\spinz\finket\irkeoq,
 \end{array}\hspace{0.75cm}
\label{eq:onepsa} \end{eqnarray}
 where $\ketvide$ is the vacuum state,
 $\opcrea\!\left(\vkond,[\spinz]_{\cZ[\vkond]}\right)$
 is the creation operator of a particle of momentum $\hbar\vkond$
 and spin component $\spinz$ on the quantization axis $\cZ[\vkond]$
 and $\debket\spinz\finket\irkeoq$ is the eigenstate
 of spin component $\spinz$ on $\cZ[\vkond]$.
 The initial state is given by:
\begin{eqnarray}\debket\etatm^{(\modspin)}_{\textup{in}}\finket= \left\{
\begin{array}{lll}\displaystyle \debket\vkondini\finket 
 &\hspace{0.5cm}\mbox{if}& \modspin=0
\vspace{0.1cm}\\ \displaystyle  \debket\vkondini\finket\!\ptens\!
 \debket\kisin\finket
 &\hspace{0.5cm}\mbox{if}& \modspin\neq 0,
 \end{array}\right.\hspace{0.75cm}
\label{eq:onepsi} \end{eqnarray}
 where $\debket\kisin\finket$ is the initial state of spin polarization
 prepared with the amplitudes $\ilkoeoq\debpscal\spinz\midpscal\kisin\finpscal$.
\\

}

\paragraph{Structure of the diffraction operator.}
{
 From (\!\!~\ref{eq:fgenera}) and (\!\!~\ref{eq:onepsi}),
 the non-normalized final state for a particle without spin is expressed by:
\begin{eqnarray} \opedif^{(0)}\!\debket\etatm^{(0)}_{\textup{in}}\finket
 \!\!= \opedif^{(0)}\!\debket\vkondini\finket\!\!=\!\!\inttrp\!\differ^3\mkond
 \,\debket\vkond\finket\!\!\debvalop\!\vkond\!\midvalopa\opedif^{(0)}
 \midvalopb\!\vkondini\!\finvalop\!\!.\hspace{0.75cm}
\label{eq:onepsf} \end{eqnarray}
 To generalize this expression to the case of a particle of non-zero spin,
 we rely on the following observation.
 For the photon, the quantization axis is in the direction
 of the momentum and the eigenvalue zero of the spin component
 is impossible \cite{LLrel}.
 Therefore, the change in the direction of the momentum of the photon
 due to diffraction causes a modification of its spin polarization
 so that this impossibility of the eigenvalue zero is preserved.
 More generally, we assume that for any particle, the momentum exchange
 with the diaphragm causes a specific change in spin polarization.
 
 The change in polarization corresponds to a rearrangement of the spin
 component wave functions and therefore results from the action of a
 unitary rotation operator.
 So we are led to assume that if the measurement of the momentum
 of the detected particle gives the result $\hbar\vkond$ then
 the probabilities of the results of a simultaneous measurement of the
 spin component correspond to a polarization state which depends
 on $\vkond$ in the form:
\begin{eqnarray} \debket\kisoutk\finket
 =\oprotsp^{(\modspin)}[\,\eulera(\vkond),\eulerb(\vkond),\eulerc(\vkond)\,]
 \,\debket\kisin\finket\!,\hspace{0.75cm}
\label{eq:rotinid} \end{eqnarray}
 where $\oprotsp^{(\modspin)}[\,\eulera(\vkond),\eulerb(\vkond),
 \eulerc(\vkond)\,]$ is the {\it operator of the spin rotation
 associated with the momentum transfer $\hbar\vkondini\rightarrow\hbar\vkond$}.
 The state $\debket\kisoutk\finket$ is in some way the "conditional state"
 of polarization associated with the momentum eigenstate $\debket\vkond\finket$.
 The Euler angles $\eulea_{\entj}(\vkond)$ are defined with respect to
 the quantization axis $\cZ[\vkondini]$ and are three parameters of the model.
 They are functions of $\vkond$, not known a priori.
 They also depend on $\vkondini$ and possibly on other parameters
 as for example the spin of the particle: $\eulea_{\entj}(\vkond)\equiv
 \eulea_{\entj}^{\vkondini,\modspin,...}(\vkond)$.

 An additional assumption is needed to generalize (\!\!~\ref{eq:onepsf}).
 For a spinless particle, the position and momentum wave functions are Fourier
 transforms of each other. In the case of diffraction with a diaphragm,
 the shape of the final momentum distribution is therefore determined
 by the shape of the aperture.
 We assume that this determination is the same if the particle has spin,
 so that the final momentum distribution of a particle with spin
 is the same as that of a spinless particle which would have the same energy.
 There do not seem to be any experimental facts invalidating this assumption.
 
 The easiest way to generalize (\!\!~\ref{eq:onepsf}) taking into account
 (\!\!~\ref{eq:onepsi}), (\!\!~\ref{eq:rotinid}) and the additional assumption
 above is to express the action of $\opedif^{(\modspin)}$ on the initial state
 in the following form
 (we use the notation $\oprotsp^{(\modspin)}(\vkond)$ instead of
 $\oprotsp^{(\modspin)}[\,\eulera(\vkond),\eulerb(\vkond),\eulerc(\vkond)\,]$
 for simplicity and we insert the identity operator
 $\sum_{\spinz} \debket\spinz\finket\irkeoq\;\ilkeoq\debbra\spinz\finbra$):
\begin{eqnarray}\begin{array}{l}\displaystyle \modspin \neq 0: \;\;\;
 \opedif^{(\modspin)}\debket\etatm^{(\modspin)}_{\textup{in}}\finket
 =\;\opedif^{(\modspin)}\left(\,\debket\vkondini\finket\!\! \ptens\!
 \debket\kisin\finket\right)
\vspace{0.2cm}\\ \displaystyle \hspace{1cm}
 =\!\inttrp\!\differ^3\mkond\debket\vkond\finket\!\debvalop\vkond\midvalopa
 \opedif^{(0)}\midvalopb\vkondini\finvalop\!\ptens\!\debket\kisoutk\finket
\vspace{0.2cm}\\ \displaystyle \hspace{1cm}
 =\!\inttrp\!\differ^3\mkond\debket\vkond\finket\!
 \debvalop\vkond\midvalopa\opedif^{(0)}\midvalopb\vkondini\finvalop
\\ \displaystyle \hspace{1.5cm}
 \ptens \sum_{\spinz}\debket\spinz\finket\irkeoq\;\ilkeoq\debvalop\spinz
 \midvalopa\oprotsp^{(\modspin)}(\vkond)\midvalopb\kisin\!\finvalop\!.
\end{array}\hspace{0.75cm}
\label{eq:onepsg} \end{eqnarray}
 From (\!\!~\ref{eq:fgenera}), (\!\!~\ref{eq:onepsi}), (\!\!~\ref{eq:onepsf})
 and (\!\!~\ref{eq:onepsg}), the final state is a linear combination of the
 momentum and spin component eigenstates given by (\!\!~\ref{eq:onepsa}) and
 the diffraction operator is:
\begin{eqnarray}\!\opedif^{(\modspin)}=\left\{\begin{array}{cl}\displaystyle
 \opedif^{(0)}&\;\mbox{ if }\modspin=0 \vspace{0.1cm}\\ \displaystyle
 \inttrp\!\differ^3\mkond\,\debket\vkond\finket\!\!\debbra\vkond\finbra
 \opedif^{(0)}\!\!\ptens\!\oprotsp^{(\modspin)}(\vkond)&\;
 \mbox{ if }\modspin\neq 0.
 \end{array}\right.\hspace{0.75cm}
\label{eq:onepsp} \end{eqnarray}
 The operator $\opedif^{(0)}$ will be called {\it "momentum part"}
 of the diffraction operator $\opedif^{(\modspin)}$.
\vspace{0.2cm}
}

\paragraph{General expressions of the final amplitudes and probabilities.}
{
 From (\!\!~\ref{eq:rotinid}) and since $\oprotsp^{(\modspin)}(\vkond)$
 is unitary:
\begin{eqnarray} \debpscal\kisoutk\midpscal\kisoutk\finpscal
 =\debpscal\kisin\midpscal\kisin\finpscal=1. \hspace{0.5cm}
\label{eq:rotuncn} \end{eqnarray}
 From (\!\!~\ref{eq:fgenerb}) into which we substitute
 (\!\!~\ref{eq:onepsf}) (if $\modspin=0$) or (\!\!~\ref{eq:onepsg})
 (if $\modspin\neq 0$) and given (\!\!~\ref{eq:rotuncn}),
 we find that the normalization factor is independent of the spin:
\begin{eqnarray} \forall\modspin:\hspace{0.25cm}\norm^{(\modspin)}\equiv\norm
 =\!\int\!\differ^3\mkond \,\left|\debvalop\vkond\midvalopa\opedif^{(0)}
 \midvalopb\vkondini\finvalop\right|^2. \hspace{0.75cm}
\label{eq:fgenerw} \end{eqnarray}

 If $\modspin=0$, the probability amplitude to detect the particle
 with momentum $\hbar\vkond$ is obtained by substituting (\!\!~\ref{eq:onepsf})
 into (\!\!~\ref{eq:fgenera}). Given (\!\!~\ref{eq:onepsi}) and
 (\!\!~\ref{eq:fgenerw}), this leads to:
\begin{eqnarray} \debpscal\vkond\midpscal\etatm^{(0)}_{\textup{out}}\finpscal=
 \norm^{-1/2}\debvalop\vkond\midvalopa\opedif^{(0)}\midvalopb\vkondini
 \finvalop\!. \hspace{0.75cm}
\label{eq:finstc} \end{eqnarray}
 The p.d.f. to detect the particle with momentum $\hbar\vkond$ is:
\begin{eqnarray} \pdfp^{(0)}_{\vavkond}(\vkond)=\left|\debpscal\!\vkond\!
 \midpscal\etatm^{(0)}_{\textup{out}}\finpscal\right|^{2}.
\label{eq:fgeners} \end{eqnarray}

 If $\modspin\neq 0$, the probability amplitude to detect the particle
 with momentum $\hbar\vkond$ and spin component $\spinz$
 on the $\cZ[\vkond]$ axis is obtained by substituting (\!\!~\ref{eq:onepsg})
 into (\!\!~\ref{eq:fgenera}). Given (\!\!~\ref{eq:fgenerw}) and
 (\!\!~\ref{eq:finstc}), this leads to:
\begin{eqnarray}\begin{array}{l}\displaystyle
 \left(\!\debbra\vkond\finbra\!\!\ptens\!\!\ilkeoq\debbra\spinz\finbra\!\left)
 \debket\etatm^{(\modspin)}_{\textup{out}}\finket\!\!\right.\right.
\vspace{0.2cm}\\ \displaystyle \hspace{2.5cm}
 \!=\!\debpscal\!\vkond\!\midpscal\etatm^{(0)}_{\textup{out}}\finpscal\!
 \ilkeoq\debpscal\!\spinz\!\midpscal\kisoutk\finpscal\!\!.
\end{array} \hspace{0.75cm}
\label{eq:fgenert} \end{eqnarray}
 The joint probability function to detect the particle
 with momentum $\hbar\vkond$ and spin component $\spinz$ on the
 $\cZ[\vkond]$ axis is expressed, according to the definition of the conditional
 probability and from (\!\!~\ref{eq:fgenert}), by:
\begin{eqnarray}\!\begin{array}{l}\displaystyle
 \pdfpj^{(\modspin)}_{\vavkond,\vaspinz}
 \!\left(\vkond,[\spinz]_{\cZ[\vkond]}\right)
 \!=\!\pdfp^{(\modspin)}_{\vavkond}(\vkond)\,
 \probcon_{\vaspinzc}^{(\modspin)}\left([\spinz]_{\cZ[\vkond]}\right)
\vspace{0.2cm} \\ \displaystyle \hspace{2cm}
 =\left|\!\debpscal\!\vkond\!\midpscal
 \etatm^{(0)}_{\textup{out}}\finpscal\!\right|^{2}
 \,\left|\ilkeoq\debpscal\spinz\midpscal\kisoutk\finpscal\right|^2\!\!,
 \end{array} \hspace{0.7cm}
\label{eq:fgenerm} \end{eqnarray}
 where $\pdfp^{(\modspin)}_{\vavkond}(\vkond)$ is the p.d.f. to detect,
 without polarization measurement, the particle with momentum $\hbar\vkond$
 and $\probcon_{\vaspinzc}^{(\modspin)}\left([\spinz]_{\cZ[\vkond]}\right)$
 is the conditional probability to detect the particle with spin component
 $\spinz$ on the $\cZ[\vkond]$ axis if its momentum is $\hbar\vkond$.
\vspace{0.15cm}

 If $\modspin\neq 0$, $\pdfp^{(\modspin)}_{\vavkond}(\vkond)$ is
 the marginal p.d.f. obtained by summing (\!\!~\ref{eq:fgenerm}) over $\spinz$.
 Given (\!\!~\ref{eq:rotuncn}) and (\!\!~\ref{eq:fgeners}), this leads to
 $\pdfp^{(\modspin)}_{\vavkond}(\vkond)=\pdfp^{(0)}_{\vavkond}(\vkond)$.
 Hence, given (\!\!~\ref{eq:finstc}) and (\!\!~\ref{eq:fgeners}):
\begin{eqnarray}\forall\modspin:\hspace{0.25cm}
 \pdfp^{(\modspin)}_{\vavkond}(\vkond)\equiv\pdfp^{}_{\vavkond}(\vkond)
 =\norm^{-1}\!\left|\debvalop\vkond\midvalopa
 \opedif^{(0)}\!\midvalopb\vkondini\finvalop\right|^{2},\hspace{0.75cm}
\label{eq:fgenerq} \end{eqnarray}
 which expresses that the momentum p.d.f.  of the detected particle
 without polarization measurement is independent from its spin
 and initial polarization.
 Moreover, substituting (\!\!~\ref{eq:finstc}) into (\!\!~\ref{eq:fgenerm})
 and given (\!\!~\ref{eq:fgenerq}), we get:
\begin{eqnarray} \probcon_{\vaspinzc}^{(\modspin)}
 \left([\spinz]_{\cZ[\vkond]}\right)
 =\left|\ilkeoq\debpscal\spinz\midpscal\kisoutk\finpscal\right|^2.
\hspace{0.75cm}
\label{eq:fgenerr} \end{eqnarray}

 The experimentaly accessible quantities are those given by
 (\!\!~\ref{eq:fgenerq}) and (\!\!~\ref{eq:fgenerr}).
 To calculate them, we therefore need to express the matrix elements
 $\debvalop\vkond\midvalopa\opedif^{(0)}\midvalopb \vkondini\finvalop$
 and the amplitudes $\ilkeoq\debpscal\spinz\midpscal\kisoutk\finpscal$.
 This is the subject of the next two subsections.

}

}

\subsection{Momentum part of the diffraction operator}
{
\label{sub:projini}

 In this subsection, we first deal with the case of non-relativistic particles.
 We will then show that the developed formalism
 can be transposed to the case of photons.
\\
\paragraph{Position measurement and the Huygens-Fresnel principle.}
{
 At the time $\tmes$ of the position measurement, the position wave function
 of the particle undergoes a localization at the aperture
 of the diaphragm (postulate of wave function reduction).
 During this temporary localization, the transverse coordinates of the particle
 correspond to the aperture and the longitudinal coordinate is equal or close to
 $\cz=0$ since the particle then crosses the plane of the diaphragm.
 The position measurement is therefore a measurement
 of the {\it three} spatial coordinates.
\\

 The measurement of the transverse coordinates is associated with the projector:
\begin{eqnarray}
 \Projtran^{\ouvas}\equiv\intdbl_{\ouvas}\differ\cx\differ\cy\,
 \debket\cx\cy\finket\!\!\debbra\cx\cy\finbra,
\label{eq:proxy}\end{eqnarray}
 where $\ouvas$ is the aperture.
 Then, the easiest way to describe the measurement of $\cz$ is
 to use a projector of the form:
\begin{eqnarray} \Projlong^{\Delta\cz}\;\equiv\;
 \int_{-\Delta\cz/2}^{+\Delta\cz/2}\differ\cz\,
 \debket\cz\finket\!\!\debbra\cz\finbra,
\label{eq:proz}\end{eqnarray}
 where the width $\Delta\cz$ of the interval $[-\Delta\cz/2,+\Delta\cz/2]$
 is a parameter of the model whose value is not known a priori.
 Finally, the measurement of the three coordinates $(\cx,\cy,\cz)$ is assumed
 to be associated with the projector:
\begin{eqnarray}
 \Proj^{\ouvas,\Delta\cz}\;\equiv\;
 \Projtran^{\ouvas}\otimes\Projlong^{\Delta\cz}.
\label{eq:proxyz}\end{eqnarray}

 Since the aperture $\ouvas$ is a 2D surface, we should have in principle:
 $\Delta\cz=0$. But the integral of the right-hand side of (\!\!~\ref{eq:proz})
 is zero in this case. Suppose then that $\Delta\cz\neq 0$.
 From (\!\!~\ref{eq:proxyz}), we have:
 $\Proj^{\ouvas,\Delta\cz}\!\debket\vkondini\finket\!=\Projtran^{\ouvas}
 \debket\kox\,\koy\finket\!\ptens\Projlong^{\Delta\cz}\debket\koz\finket$.
 Therefore, from (\!\!~\ref{eq:proz}), the p.d.f. corresponding to
 the probability of finding a result within the interval $[\cz,\cz+\differ\cz]$
 when measuring the longitudinal coordinate is proportional to:
\begin{eqnarray}\!\!\begin{array}{l}\displaystyle
 \left|\!\debvalop\!\cz\!\midvalopa\Projlong^{\Delta\cz}\midvalopb
 \!\koz\!\finvalop\!\right|^2
\vspace{0.2cm}\\ \displaystyle \hspace{2cm}
 = \left\{\!\begin{array}{cll}
 (2\pi)^{-1}&\;\mbox{if}&\!\cz\in[\,-\Delta\cz/2,+\Delta\cz/2\,]\vspace{0.1cm}\\
 0               &\;\mbox{if}&\!\cz\notin[\,-\Delta\cz/2,+\Delta\cz/2\,].
 \end{array} \right. 
\end{array} \hspace{0.5cm}
\label{eq:probpjr} \end{eqnarray}
 If $\Delta\cz$ is small, the action of $\Proj^{\ouvas,\Delta\cz}$ localizes
 the probability of presence of the particle in a narrow region around
 the wavefront at the aperture and consequently its longitudinal coordinate
 is $\cz=0$ with excellent accuracy.
 This localization of the probability of presence occurs at time $\tmes$.
 Therefore, at any time $\temps>\tmes$, the diffracted wave  has been emitted
 from a volume including the wavefront at the aperture and its close vicinity.
 We are then close to a situation consistent with the Huygens-Fresnel principle.
 Perfect compatibility would therefore be obtained if $\Delta\cz = 0$.
 But in this case, it is not possible to obtain a p.d.f. from the function
 expressed by (\!\!~\ref{eq:probpjr}) because it is zero everywhere except
 at $\cz=0$ where its value is finite.
 However, if the value at $\cz=0$ were infinite,
 we would obtain a p.d.f. equal to the Dirac distribution $\delta(z)$.
 Thus, given the good agreement between the measurements performed so far
 and the predictions of the classical theories based on the Huygens-Fresnel
 principle, this is worth looking for a way to treat this limit case.
 Fortunately, it turns out that this is possible provided, however,
 that the notion of projector is generalized.
\\

}

\paragraph{Position filtering operator. "Multi-wavefronts" Huygens-Fresnel principle.}
{
\label{par:posfilt}

 If $\Delta\cz=0$, a p.d.f. equal to $\delta(\cz)$ can be obtained
 by using, instead of the projector (\!\!~\ref{eq:proz}),
 a {\it filtering operator} $\opredulong^{\Delta\cz}$ defined as:
\begin{eqnarray} \opredulong^{\Delta\cz}\equiv
 \int\differ\cs\;\sqrt{\fredDzl(\cs)}\;
 \debket\cs\finket\!\!\debbra\cs\finbra,\hspace{0.5cm}
\label{eq:projods} \end{eqnarray}
 where $\fredDzl(\cs)$ is a positive function normalized to 1,
 such that its integral outside the interval
 $[\,-\Delta\cz/2,+\Delta\cz/2\,]$ is negligible, and such that:
\begin{equation} \lim_{\Delta\cz\rightarrow 0}\fredDzl(\cz)=\delta(\cz).
\label{eq:projodt} \end{equation}
 From (\!\!~\ref{eq:projods}):
\begin{eqnarray}\begin{array}{rcl}\displaystyle
 \left|\!\debvalop\!\cz\!\midvalopa\opredulong^{\Delta\cz}\midvalopb
 \!\koz\!\finvalop\!\right|^2\!\!
 &=&\left|\!\debpscal\cs\midpscal\etatiz\finpscal\!\right|^2\,\fredDzl(\cz)
\vspace{0.2cm}\\ \displaystyle \hspace{2.75cm}
 &=&(2\pi)^{-1}\,\fredDzl(\cz).
\end{array} \hspace{0.75cm}
\label{eq:projodw} \end{eqnarray}
 Therefore, given (\!\!~\ref{eq:projodt}), if $\Delta\cs=0$,
 $\left|\!\debvalop\!\cz\!\midvalopa
 \opredulong^{\Delta\cz}\midvalopb\!\koz\!\finvalop\!\right|^2$ is defined
 and proportional to $\delta(\cs)$.
 This allows to obtain a p.d.f. equal to $\delta(\cs)$ after normalization.
 
 However, the problem is not completely solved because,
 from (\!\!~\ref{eq:projods}) and (\!\!~\ref{eq:projodt}),
 $\opredulong^{\Delta\cz}$ is not defined if $\Delta\cz=0$
 since the square root of $\delta(\cz)$ is not defined.
 So we are in a way compelled to assume that $\Delta\cz$ is not zero
 (but possibly close to zero, so that the p.d.f. can then be expressed
 with a good approximation by the Dirac distribution).
 This implies reviewing the question of the connection between diffraction
 and the Huygens-Fresnel principle.
 The case $\Delta\cz=0$ corresponds to the Kirchhoff integral where
 a "single-wavefront" Huygens-Fresnel principle is applied:
 the wavelets contributing to the diffracted wave are emitted from
 one wavefront located at the aperture.
 The case $\Delta\cz>0$, suggested by the quantum approach,
 would then correspond to a "multi-wavefronts" Huygens-Fresnel principle
 where several neighboring wavefronts contribute with different weights
 whose distribution is the function $\fredDzl(\cz)$.

 Moreover, from the first equality of (\!\!~\ref{eq:projodw}), $\fredDzl(\cz)$
 can also be interpreted as the weight with which the filtering operator selects
 the result $\cz$ from the value at $\cz$ of the position wave function
 in the initial state $\debket\etatiz\finket$.
 This weight, as a function of $\cz$, will be called
 {\it longitudinal position filtering function}.

 For the transverse coordinates, the projector (\!\!~\ref{eq:proxy})
 can be replaced by the filtering operator:
\begin{eqnarray} \opredutran^{\ouvas}\equiv\intdbl\differ\cx\differ\cy
 \;\sqrt{\fredAt(\cx,\cy)}\;\debket\cx\cy\finket\!\!\debbra\cx\cy\finbra,
 \hspace{0.75cm}
\label{eq:redxy}\end{eqnarray}
 where $\fredAt(\cx,\cy)$ is the {\it transverse position filtering function}.
 It can be considered that the transmission of the incident wave is the same over
 the entire area of the aperture so that this function corresponds to a
 {\it uniform} filtering which truncates the wave function. Hence:
\begin{equation} \fredAt(\cx,\cy)\;=\;\surfouva^{-1}\times
 \left\{ \begin{array}{cll}
 1&\mbox{ if }& (\cx,\cy)\in\ouvas \\
 0&\mbox{ if }& (\cx,\cy) \notin\ouvas, \end{array} \right.
\label{eq:tronqfod} \end{equation}
 where $\surfouva$ is the area of $\ouvas$. From (\!\!~\ref{eq:proxy}),
 (\!\!~\ref{eq:redxy}) and (\!\!~\ref{eq:tronqfod}):
 $\opredutran^{\ouvas}=\surfouva^{-1/2}\,\Projtran^{\ouvas}$,
 so that the action of the two operators leads to the same state
 after normalization.
 More generally, any projector is equivalent to a uniform filtering operator.

 The filtering operator allows to consider the case
 of a non-uniform filtering. In particular, the longitudinal filtering
 could be non-uniform contrary to the transverse filtering
 because the aperture is limited by a material edge in the transverse plane
 whereas there are no edges along the longitudinal direction.
 The longitudinal filtering function could then be a continuous function
 forming a peak centered around $\cz=0$ and of width $\Delta\cz$.
 The precise shape of the filtering function is part of the
 assumptions of the model. This shape may matter
 if $\Delta\cz$ is large but probably not if $\Delta\cz$
 is close to zero because the p.d.f. is then close to the Dirac distribution.
\\

 Finally, given ($\!\!$~\ref{eq:projods}) and ($\!\!$~\ref{eq:redxy}),
 we replace the projector $\Proj^{\ouvas,\Delta\cz}$ defined in
 ($\!\!$~\ref{eq:proxyz}) by the filtering operator:
\begin{eqnarray}\begin{array}{c}\displaystyle
 \opredu^{\ouvas,\Delta\cz}\equiv
 \opredutran^{\ouvas}\ptens\opredulong^{\Delta\cz}=\!\int\!\differ^3\rvec\,
 \sqrt{\fred^{\ouvas,\Delta\cz}(\rvec)}
 \,\debket\rvec\finket\!\!\debbra\rvec\finbra,
\vspace{0.2cm}\\ \displaystyle
 \fred^{\ouvas,\Delta\cz}(\rvec)\;\equiv\;
 \fredAt(\cx,\cy)\;\fredDzl(\cz).
\end{array}\hspace{0.75cm}
\label{eq:tronqfoc} \end{eqnarray}
 The volume $\ouvas\times[-\Delta\cz/2,+\Delta\cz/2]$
 of transverse section $\ouvas$ and length $\Delta\cz$,
 centered at the origin O is called {\it 3D aperture}. 
 The 3D aperture can be defined as the region where the position wave function
 of the particle is temporarily localized during the position measurement.
 The aperture $\ouvas$ and the interval $[-\Delta\cz/2,+\Delta\cz/2]$
 are respectively called {\it transverse 2D aperture} and
 {\it longitudinal 1D aperture} (Fig. $\!\!$~\ref{fig:ouvgenz}).

\begin{figure}[h]
\centering
\begin{minipage}[t]{8.5cm}
\resizebox{1.2\textwidth}{!}{
\includegraphics{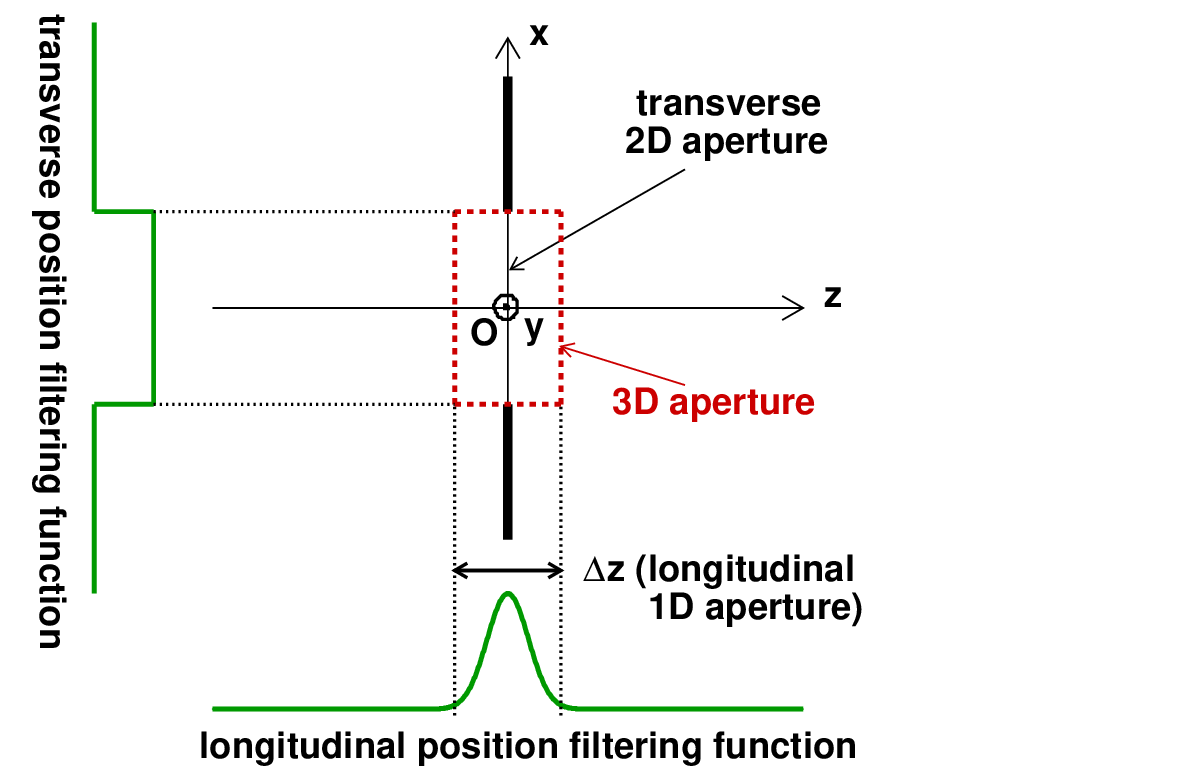}}
\caption{
 Example of 3D aperture (section in the $(\Ox,\Oz)$ plane) with the
 corresponding transverse and longitudinal position filtering functions.
}
 \label{fig:ouvgenz}
 \end{minipage}
\end{figure}

 In the case of a uniform filtering, the aperture is the region
 where the filtering function is non-zero. In the case of a non-uniform
 filtering, the filtering function can be non-zero everywhere
 (for example if it is a Gaussian). We are then led to define more generally
 the aperture as the region outside of which the integral
 of the filtering function is negligible.

 In ($\!\!$~\ref{eq:tronqfoc}), $\Delta\cz$ does not depend on $\cx$ and $\cy$,
 which is an implicit assumption in the definition ($\!\!$~\ref{eq:projods}).
 More generally, the {\it position filtering operator} is defined by:
\begin{eqnarray}\opredu^{\ouva}=\inttrp\!\differ^3\rmod\,
 \sqrt{\fredA(\rvec)}\;\debket\rvec\finket\!\!\debbra\rvec\finbra,\hspace{0.75cm}
\label{eq:tronqfob} \end{eqnarray}
 where $\ouva$ is the 3D aperture whose shape can be assumed
 to be more or less complicated and $\fredA(\rvec)$ is the
 {\it position filtering function} whose expression can be assumed
 to be different from  a product of the form (\!\!~\ref{eq:tronqfoc}).
\\

}

\paragraph{Need to consider kinematics.}
{
 From (\!\!~\ref{eq:tronqfob}), $\left|\debvalop\rvec\midvalopa
 \opredu^{\ouva}\midvalopb\vkondini\finvalop\right|^2$ is proportional to
 $\fredA(\rvec)$. So the state $\opredu^{\ouva}\debket\vkondini\finket$
 is associated with the momentum p.d.f. of the particle just after its
 localization at the aperture, when it is about to move away from the diaphragm.
 Moreover, from (\!\!~\ref{eq:fgenerq}), the state
 $\opedif^{(0)}\debket\vkondini\finket$ corresponds to the momentum p.d.f.
 $\pdfp_{\vavkond}(\vkond)$ of the particle detected beyond the diaphragm.
 Since the particle is free after its passage through the aperture,
 its momentum is conserved until its detection,
 which suggests that $\opedif^{(0)}$ is nothing other than $\opredu^{\ouva}$.
 However, this cannot be the case for the following reason.
 From (\!\!~\ref{eq:tronqfob}), the momentum wave function of the state
 $\opredu^{\ouva}\debket\vkondini\finket$ is expressed by:
\begin{eqnarray}\debvalop\vkond\midvalopa\opredu^{\ouva}\midvalopb
 \vkondini\finvalop\!=(2\pi)^{-3/2}\;
 \jfour^{\ouva}(\vkond-\vkondini),\hspace{0.75cm}
\label{eq:elmafsq} \end{eqnarray}
 where $\jfour^{\ouva}(\vkond-\vkondini)$ is the Fourier transform
 of the square root of the position filtering function:
\begin{eqnarray}\begin{array}{l} \displaystyle
 \jfour^{\ouva}(\vkond-\vkondini)\,\equiv\,(2\pi)^{-3/2}
\vspace{0.1cm} \\ \displaystyle \hspace{1.75cm} \times
 \!\!\inttrp\!\differ^3\rmod\;\sqrt{\fredA(\rvec)}\;
 \exp\left[\,-\icmp(\vkond-\vkondini)\,\pscal\,\rvec\,\right].
 \end{array} \hspace{0.75cm}
\label{eq:nrspinb} \end{eqnarray}
 If $\opedif^{(0)}$ is equal to $\opredu^{\ouva}$,
 the p.d.f. $\pdfp_{\vavkond}(\vkond)$ is obtained by
 substituting (\!\!~\ref{eq:elmafsq}) into (\!\!~\ref{eq:fgenerq}).
 Then, the widths $\Delta\kx$, $\Delta\ky$ and $\Delta\kz$
 of this p.d.f. are those of the distribution associated with the Fourier
 transform $\jfour^{\ouva}(\vkond-\vkondini)$ and are therefore related to
 the widths $\Delta\cx$, $\Delta\cy$ and $\Delta\cz$ of the 3D aperture
 by the uncertainty relations.
 However, if $\Delta\cx$ for example is small enough,
 the relation $\Delta\cx\Delta\kx\gtrsim 1$ implies that
 $\Delta\kx$ can be sufficiently large so that $|\kx|>\mkondini$
 with non-zero probability and therefore
 the relation (\!\!~\ref{eq:contrax}) is not satisfied in such a case.
 But this is not possible because (\!\!~\ref{eq:contrax}) results from
 kinematics and is moreover confirmed by experiment.
 This issue comes from the fact that the position wave function
 of the state $\opredu^{\ouva}\debket\vkondini\finket$ is localized
 in the 3D aperture $\ouva$ and that consequently its momentum wave function is
 spread out, which results in a spreading of the distribution
 of the momentum modulus and therefore of the energy.
 For (\!\!~\ref{eq:contrax}) to be satisfied, we are led to assume that
 $\opedif^{(0)}$ is not simply equal to $\opredu^{\ouva}$
 but is rather of the form:
\begin{eqnarray}\opedif^{(0)}=\,\opredu^{\mkondini}\;\opredu^{\ouva},
\label{eq:opdifra} \end{eqnarray}
 where $\opredu^{\mkondini}$ is an {\it energy-momentum filtering operator}
 whose role is to act on the state $\opredu^{\ouva}\debket\vkondini\finket$
 - which is then a {\it localized transitional state} -
 to obtain a final state of same energy as that of the initial state.
\\

}

\paragraph{Energy-momentum filtering operator.}
{
 The filtering operator $\opredu^{\mkondini}$ must be associated
 with the domain $\ouvcin$ of the momentum space that corresponds
 to the vectors $\vkond$ compatible with kinematics. So we define,
 using an expression similar to (\!\!~\ref{eq:tronqfob}):
\begin{equation}\opredu^{\mkondini}\equiv\!\inttrp\differ^3\mkond\;
 \sqrt{\fredcin(\vkond)}\;\debket\vkond\finket\!\!\debbra\vkond\finbra,
\label{eq:transfob} \end{equation}
 where $\fredcin(\vkond)$ is a {\it momentum-energy filtering function}
 which must represent the weight with which the filtering operator
 selects the result $\vkond$ from the value at $\vkond$ of the
 momentum wave function in the localized transitional state
 $\opredu^{\ouva}\debket\vkondini\finket$.
 From (\!\!~\ref{eq:contrax}) and (\!\!~\ref{eq:contray}),
 we are led to assume that this function is of the form:
\begin{eqnarray}\fredcin(\vkond)\equiv\consn\;
 \gaussmk\!\left(|\vkond|-\mkondini\right)\;\delta_{1\,\sgn[\kz]},\hspace{0.5cm}
\label{eq:transfoc} \end{eqnarray}
 where $\consn$ is a normalization constant that will be calculated below,
 $\gaussmk\left(|\vkond|-\mkondini\right)$ is a function of the modulus
 of $\vkond$ forming a peak centered at $|\vkond|=\mkondini$
 and of width $\Delta\mkond$ close to zero
 (in accordance with (\!\!~\ref{eq:contrax})) and
 the Kronecker delta $\delta_{1\,\sgn[\kz]}$ ensures that
 $\fredcin(\vkond)$ is zero if $\kz\leq 0$
 (in accordance with (\!\!~\ref{eq:contray})).
 From (\!\!~\ref{eq:transfoc}), using the spherical coordinates,
 the normalization to 1 of $\fredcin(\vkond)$ is expressed by:
\begin{eqnarray}\!\!
 1=\consn\!\!\int_{0}^{\infty}\!\!\!\!\!\!\differ\mkond\;\mkond^2\,
 \gaussmk\!\left(\mkond\!-\!\mkondini\right)
 \!\!\int_{0}^{\pi}\!\!\!\!\!\differ\theta\,\sin\!\theta\,
 \delta_{1\,\sgn[\cos\theta]}
 \!\int_{0}^{2\pi}\!\!\!\!\!\!\differ\phi.\hspace{0.75cm} 
\label{eq:transfod} \end{eqnarray}
 Since $\Delta\mkond$ is close to zero, we can replace
 $\gaussmk\left(\mkond-\mkondini\right)$ by $\delta(\mkond-\mkondini)$ in
 the integral over $\mkond$ whose value is therefore close to ${\mkondini}^2$.
 Then, (\!\!~\ref{eq:transfod}) implies:
 $\consn\simeq{\mkondini}^{\!-2}\,(2\pi)^{-1}$.
 Substituting (\!\!~\ref{eq:transfoc}) with this value of $\consn$
 into (\!\!~\ref{eq:transfob}), we get:
\begin{eqnarray}\begin{array}{l}\displaystyle
 \opredu^{\mkondini}\simeq\,(2\pi)^{-1/2}\,\mkondini^{-1}
 \vspace{0.2cm}\\ \displaystyle \hspace{1cm} \times\!
 \!\inttrp\!\differ^3\mkond\;\sqrt{\gaussmk\!\left(|\vkond|-\mkondini\right)}
 \;\delta_{1\,\sgn[\kz]}\;\debket\vkond\finket\!\!\debbra\vkond\finbra.
\end{array} \hspace{0.75cm}
\label{eq:transfon} \end{eqnarray}
 We can interpret $\opredu^{\mkondini}$ as an operator which represents
 an energy-momentum measurement including a measurement of the momentum modulus
 (in other words of the energy) giving the result $\hbar\mkondini$
 with near certainty and a measurement of the momentum longitudinal component
 giving the result $\hbar\kz>0$.
\vspace{0.2cm}
}

\paragraph{Matrix element of the momentum part of the diffraction operator.}
{
 Substituting (\!\!~\ref{eq:tronqfob}) - in which we insert
 the identity operator $\inttrp\differ^3\mkond\,\debket\vkond\finket
 \!\debbra\vkond\finbra$ after $\debket\rvec\finket\!\debbra\rvec\finbra$ -
 and (\!\!~\ref{eq:transfon}) into (\!\!~\ref{eq:opdifra}), and given
 (\!\!~\ref{eq:nrspinb}), we obtain:
\begin{eqnarray}\!\!\!\!\begin{array}{l} \displaystyle
 \opedif^{(0)}\!\simeq\,(2\pi)^{-2}\,\mkondini^{-1}
 \!\!\inttrp\!\differ^3\mkond\,\sqrt{\gaussmk\!\left(|\vkond|\!-\!\mkondini\right)}
 \;\delta_{1\,\sgn[\kz]}
\vspace{0.1cm} \\ \displaystyle \hspace{2.75cm} \times
 \!\!\inttrp\!\differ^3\mkond'\;\jfour^{\ouva}(\vkond-\vkond')
 \;\debket\vkond\finket\!\!\debbra\vkond'\finbra.
 \end{array} \hspace{0.65cm}
\label{eq:tronqfoe} \end{eqnarray}
 Hence, instead of (\!\!~\ref{eq:elmafsq}):
\begin{eqnarray}\begin{array}{l} \displaystyle
 \debvalop\vkond\midvalopa\opedif^{(0)}\midvalopb\vkondini\finvalop
 \simeq(2\pi)^{-2}\,\mkondini^{-1}
\vspace{0.2cm} \\ \displaystyle \hspace{1.5cm} \times
 \sqrt{\gaussmk\!\left(|\vkond|-\mkondini\right)}\;\delta_{1\,\sgn[\kz]}
 \;\jfour^{\ouva}(\vkond-\vkondini).
\end{array} \hspace{0.75cm}
\label{eq:opdispa} \end{eqnarray}
\\

}

\paragraph{Photons.}
{
 A position filtering operator of the form ($\!\!$~\ref{eq:tronqfob}),
 where the projector $\debket\rvec\finket\!\!\debbra\rvec\finbra$ is involved,
 cannot be used for the photon because the localized photon states are eigenstates
 of a {\it photon position operator} different from the position observable of
 the non-relativistic case.
 Several problems were encountered and then finally resolved to construct
 this photon position operator and, more generally, to elaborate
 a true quantum mechanics of the photon
 \cite{NeWi,BiBi,Sipe,Hacoc,HaBay,SmRay,Brod,HaDeb,BaMos,MaHaw}.
 The localized photon states are biorthogonal \cite{Brod}
 with a specific scalar product \cite{HaDeb} and it follows that the
 appropriate operator to replace the projector
 $\debket\rvec\finket\!\!\debbra\rvec\finbra$ in the photon case is
 $\oppotvcre(\rvec,\temps)\ketvide\!\pscal\!\bravide\opchelani(\rvec,\temps)$,
 where $\oppotvpm(\rvec,\temps)$ and $\opchelpm(\rvec,\temps)$ are the
 positive and negative frequency field operators of the transverse
 vector potential and electric field.
 These field operators are given by \cite{CTDRG}:
\begin{eqnarray}\begin{array}{l} \displaystyle
 \oppotvcre(\rvec,\temps)=\left[\oppotvani(\rvec,\temps)\right]^{\dagger}=
 \sqrt{\frac{\hbar}{2\epso}}\;(2\pi)^{-3/2}
\vspace{0.1cm}\\ \displaystyle \hspace{0.5cm}\times 
 \!\!\inttrp\!\frac{\differ^3\mkond}{\sqrt{\freq}}\sum_{\ilinp=x,y}\!
 \!\exp\left[\,\icmp\left(\freq\temps-\vkond\pscal\rvec\right)\,\right]
 \veclinp^{(\ilinp)}_{\vkond}\;\opcrea(\vkond,\ilinp[\vkond]),
\vspace{0.25cm}\\ \displaystyle \hspace{0.1cm}
 \opchelcre(\rvec,\temps)=\left[\opchelani(\rvec,\temps)\right]^{\dagger}=
 -\frac{\partial}{\partial\temps}\oppotvcre(\rvec,\temps),
 \end{array} \hspace{0.75cm}
\label{eq:locafa} \end{eqnarray}
 where $\freq=\vlum\mkond$,
 $\veclinp^{(\ilinp)}_{\vkond}$ is the unitary vector of the $\ilinp[\vkond]$
 axis of a coordinate system such that $\cz[\vkond]\parallel\vkond$
 and $\opcrea(\vkond,\ilinp[\vkond])$ is the creation operator
 of a photon of momentum $\hbar\vkond$ and linearly polarized in the direction
 of the $\ilinp[\vkond]$ axis.
 Similarly to (\!\!~\ref{eq:onepsa}), we have:
\begin{eqnarray}\opcrea(\vkond,\ilinp[\vkond])\ketvide\;=\;
 \debket\vkond\finket\!\!\ptens\!\debket\ilinp\finket\irkeoz,\hspace{0.75cm}
\label{eq:onepsb} \end{eqnarray}
 where $\debket\ilinp\finket\irkeoz$ is the basis state
 of linear polarization in the direction of the $\ilinp[\vkond]$ axis.
 From (\!\!~\ref{eq:locafa}) and (\!\!~\ref{eq:onepsb}):
\begin{eqnarray}\!\!\begin{array}{l}\displaystyle
 \oppotvcre(\rvec,\temps)\ketvide\!\pscal\!\bravide\opchelani(\rvec,\temps)
 \,=\,\frac{\icmp\hbar}{2\epso}\,(2\pi)^{-3}
\vspace{0.15cm}\\ \displaystyle \hspace{0.3cm}\times
 \!\!\inttrp\!\!\differ^3\mkond\!\!\inttrp\!\!\differ^3\mkond'
 \sqrt{\mkond'\!/\mkond\,}
 \,\exp\{\icmp\,[(\freq\!-\!\freq')\temps\!-\!(\vkond\!-\!\vkond')\pscal\rvec]\}
\vspace{0.15cm}\\ \displaystyle \hspace{0.3cm}
 \times\debket\vkond\finket\!\!\debbra\vkond'\finbra\ptens\!\!
 \sum_{\ilinp=\cx,\cy}\sum_{\ilinp'=\cx,\cy}{\veclinp^{(\ilinp)}_{\vkond}}
 \!\pscal\,\veclinp^{(\ilinp')}_{\vkond'}\debket\ilinp\finket\irkeoz\;
 \ilkeozp\debbra\ilinp'\finbra.
\end{array} \hspace{0.75cm}
\label{eq:locafd} \end{eqnarray}
 The photon has a spin 1 and this implies that its spin projection eigenstates
 are equivalent to vectors of complex components in the basis
 $\{\veclinp^{(\cx)}_{\vkond}, \veclinp^{(\cy)}_{\vkond},
 \veclinp^{(\cz)}_{\vkond}\}$ \cite{LLQM}.
 Moreover, the basis states $\debket\ilinp\finket\irkeoz$
 are specific linear combinations of the spin projection eigenstates
 \cite{CTDRG,Mess} such that $\debket\ilinp\finket\irkeoz$ is equivalent to
 the real vector ${\veclinp^{(\ilinp)}_{\vkond}}$. Therefore:
%
 ${\veclinp^{(\ilinp)}_{\vkond}}      \pscal\,\veclinp^{(\ilinp')}_{\vkond'} =\,
 {\veclinp^{(\ilinp)}_{\vkond}}^{*}\!\pscal\,\veclinp^{(\ilinp')}_{\vkond'}
 =\ilkeoz\debpscal\ilinp\midpscal\ilinp'\finpscal\irkeozp$.
%
 So the double sum over $\ilinp$ and $\ilinp'$ in ($\!\!$~\ref{eq:locafd})
 is the product of the identity operator by itself,
 successively expressed by the closure relations of the bases
 $\{\debket\xpol\finket\irkeoz,\debket\ypol\finket\irkeoz\}$ and
 $\{\debket\xpol\finket\irkeozp,\debket\ypol\finket\irkeozp\}$.
 The action of the operator $\oppotvcre(\rvec,\temps)\ketvide\!\pscal\!
 \bravide\opchelani(\rvec,\temps)$ has therefore no effect
 on the polarization states so that we can just consider
 its restriction to the subspace of the momentum states.
 So, replacing in (\!\!~\ref{eq:tronqfob})
 $\debket\rvec\finket\!\debbra\rvec\finbra$ by the right-hand side
 of ($\!\!$~\ref{eq:locafd}) without the double sum over $\ilinp$ and $\ilinp'$
 and multiplying by the factor $-2\icmp\epso/\hbar$
 to obtain the same dimension as that of $\opredu^{\ouva}$
 (length to the power $-3/2$), we are led to assume
 that the position filtering operator for the photon is:
\begin{eqnarray}\begin{array}{l}\displaystyle
 \opredu^{\ouva}_{\textup{phot.}}(\temps)=(2\pi)^{-3}
 \!\!\inttrp\!\differ^3\rmod\;\sqrt{\fredA(\rvec)}\;
 \!\inttrp\!\!\differ^3\mkond\!\!\inttrp\!\!\differ^3\mkond'\,
\vspace{0.1cm}\\ \displaystyle \hspace{0.25cm}\times
 \,\sqrt{\mkond'\!/\mkond\,}
 \,\exp\{\icmp\,[(\freq\!-\!\freq')\temps\!-\!(\vkond\!-\!\vkond')\pscal\rvec]\}
 \debket\vkond\finket\!\!\debbra\vkond'\finbra.
\end{array} \hspace{0.75cm}
\label{eq:locafc} \end{eqnarray}
 Furthermore, we express the momentum part of the diffraction operator
 in a form similar to ($\!\!$~\ref{eq:opdifra}):
\begin{eqnarray}\opedif^{(0)}_{\textup{phot.}}(\temps)=\opredu^{\mkondini}\;
 \opredu^{\ouva}_{\textup{phot.}}(\temps). \hspace{0.75cm}
\label{eq:locaff} \end{eqnarray}
 Then, substituting ($\!\!$~\ref{eq:transfon}) and ($\!\!$~\ref{eq:locafc})
 into ($\!\!$~\ref{eq:locaff}) - and given ($\!\!$~\ref{eq:nrspinb}) -
 we finally obtain:
\begin{eqnarray}\hspace{-0.12cm}\begin{array}{l}\displaystyle
 \opedif^{(0)}_{\textup{phot.}}(\temps)\simeq(2\pi)^{-2}\mkondini^{-1}
 \!\!\!\inttrp\!\!\differ^3\mkond\,
 \sqrt{\gaussmk\!\left(|\vkond|\!-\!\mkondini\right)}\;\delta_{1\,\sgn[\kz]}
\vspace{0.15cm}\\ \displaystyle \hspace{0.1cm}\times
 \!\!\inttrp\!\!\differ^3\mkond'\sqrt{\mkond'\!/\mkond\,}\,
 \exp[\,\icmp(\freq\!-\!\freq')\temps\,]\;
 \jfour^{\ouva}(\vkond\!-\!\vkond')\debket\vkond\finket\!\!\debbra\vkond'\finbra\!.
 \end{array}\hspace{0.75cm}
\label{eq:locafb} \end{eqnarray}
 By calculating the matrix element $\debvalop\vkond\midvalopa
 \opedif^{(0)}_{\textup{phot.}}(\temps)\midvalopb\vkondini\finvalop$
 from ($\!\!$~\ref{eq:locafb}), we get an expression
 with the factor $\exp[\icmp(\freq-\freqo)\temps]$. Now, from
 ($\!\!$~\ref{eq:contrax}): $\mkond\!\simeq\!\mkondini$,
 so that $\freq\!\simeq\!\freqo$. Therefore, the matrix element in question
 does not actually depend on time and we find that its expression
 is nothing other than ($\!\!$~\ref{eq:opdispa}).
 This relation can therefore be used both for non-relativistic
 particles and for photons.
\\

}

\paragraph{Characteristics of the measurement process.}
{
 From (\!\!~\ref{eq:opdifra}) and (\!\!~\ref{eq:locaff}),
 we see that the momentum part $\opedif^{(0)}$ of the diffraction operator
 depends on the momentum modulus $\mkondini$ of the incident particle.
 Therefore, the initial state is changed by the action of an operator
 which depends on this initial state itself.
 This reflects the fact that the diaphragm and the particle form
 an inseparable system during the measurement,
 in accordance with the Copenhagen interpretation of quantum mechanics.

 Moreover, using (\!\!~\ref{eq:tronqfob}), (\!\!~\ref{eq:transfon})
 and (\!\!~\ref{eq:locafc}), we can verify that
 the product of operators in the right-hand sides of
 (\!\!~\ref{eq:opdifra}) and (\!\!~\ref{eq:locaff}) is not commutative.
 This non-commutativity imposes the order in which the operators act to create
 the final state from the initial state.
 This ordrer is related to the temporal unfolding of an irreversible process
 whose sequence is: initial state $\rightarrow$ position measurement
 ($\opredu^{\ouva}$) $\rightarrow$ localized transitional state $\rightarrow$
 energy-momentum measurement ($\opredu^{\mkondini}$) $\rightarrow$ final state
 $\rightarrow$ measurement of momentum and polarization (detectors).
 The two first measurements ($\opedif^{(0)}$) are not equivalent
 to one measurement to which the uncertainty relations apply.
 These relations are satisfied for each of the two measurements.
 Let $\Delta\cx$, $\Delta\kx$ be the uncertainties of the first measurement
 which creates the localized transitional state and $\Delta'\cx$, $\Delta'\kx$
 those of the second measurement which creates the final state.
 So $\Delta\cx$ is the width of the aperture and $\Delta'\kx$
 is the width of the distribution of $\kx$ in the final state.
 In this state, we have: $-\mkond\leq\kx\leq +\mkond$, so that
 $\Delta'\kx\lesssim 2\mkond$. Hence, because of kinematics
 (Eq. (\!\!~\ref{eq:contrax})): $\Delta'\kx\lesssim 2\mkondini$ which is finite.
 Therefore, if $\Delta\cx$ is small enough, we then have:
 $\Delta\cx\Delta'\kx\lesssim 1$ but this is not a problem because $\Delta\cx$
 is associated with the first measurement while $\Delta'\kx$ is associated with
 the second measurement. On the other hand, we have: $\Delta\cx\Delta\kx\gtrsim 1$
 and $\Delta'\cx\Delta'\kx\gtrsim 1$, where $\Delta'\cx$ corresponds to the extent
 of the diffracted wave.
 We also have the relations: $\Delta\temps\Delta\freq\gtrsim 1$ and
 $\Delta'\temps\Delta'\freq\gtrsim 1$ between the lifetimes and the widths
 in energy of the transitional state and of the final state.
 We can assume that $\Delta\temps\simeq\Delta\cz/\mvit$
 where $\mvit$ is the speed of the particle.
 Because of the Huygens-Fresnel principle, it is expected that
 $\Delta\cz\simeq 0$ (\S ~\ref{par:posfilt}). So $\Delta\temps\simeq 0$.
 Moreover, given (\!\!~\ref{eq:contrax}), we have: $\Delta'\freq\simeq 0$.
 Hence: $\Delta\temps\Delta'\freq \lesssim 1$.

}

}

\subsection{Polarization amplitudes of the detected particles}
{
\label{sub:polarp}

\paragraph{General case.}
{
 The quantization axis $\cZ[\vkond]$ belongs to a coordinate system
 $\{\cX[\vkond],\cY[\vkond],\cZ[\vkond]\}$ defined by
 $\ilinq[\vkond]=\oprotos(\Eulapre,\Eulanuy,\Eularop)\,\ilinp[\vkond]$
 ($\ilinq=\cX,\cY,\cZ$; $\ilinp=\cx,\cy,\cz$), where
 $\oprotos(\Eulapre,\Eulanuy,\Eularop)$ is a rotation whose Euler angles
 are arbitrary chosen and $\{\cx[\vkond],\cy[\vkond],\cz[\vkond]\}$
 is the coordinate system attached to the particle.
 Moreover, according to (\!\!~\ref{eq:rotkok}):
 $\ilinp[\vkond]=\oprotos(\eulapre,\eulanuy,0)\,\ilinp[\vkondini]$.
 The rotation of the eigenstates has the same Euler angle as the rotation
 of the axes because a physical system in a given eigenstate must rotate
 with the coordinate system associated with the quantization axis
 to remain in this eigenstate. Therefore:
\begin{eqnarray}\begin{array}{ccl}
 \debket\spinz\finket\irkeoqe &=&\oprotsp^{(\modspin)}
 (\Eulapre,\Eulanuy,\Eularop)\debket\spinz\finket\irkeoqz,
\vspace{0.15cm}\\ 
 \debket\spinz\finket\irkeoqz&=&\oprotsp^{(\modspin)}
 (\,\eulapre,\eulanuy,0\,)\debket\spinz\finket\irkoeoqz\,.
\end{array}\hspace{0.75cm}
\label{eq:fgenerv} \end{eqnarray}
 In the present case, where the directions of the $\ilinq[\vkond]$ axes
 are defined by the rotation $\oprotos(\Eulapre,\Eulanuy,\Eularop)$,
 the angle $\Eularop$ must be mentioned in the notation
 $\debket\spinz\finket\irkeoqe$ because $\cZ[\vkond]$ only depends on
 $\Eulapre,\Eulanuy$ whereas the rotation operator $\oprotsp^{(\modspin)}
 (\Eulapre,\Eulanuy,\Eularop)$ - so a priori the resulting state -
 also depends on $\Eularop$.

 To express the final polarization amplitudes (quantization axis $\cZ[\vkond]$)
 as a function of the initial amplitudes (quantization axis $\cZ[\vkondini]$),
 we multiply the relation (\!\!~\ref{eq:rotinid}) on the left by
 $\ilkeoqe\debbra\spinz\finbra$ and we insert the identity operator
 $\sum_{\spinz'}\!\debket\!\spinz'\!\finket\!\irkoeoqe\;
 \ilkoeoqe\debbra\!\spinz'\!\finbra$ before the ket $\debket\kisin\finket$.
 We then use (\!\!~\ref{eq:fgenerv}) and the relation:
 $\oprotsp^{(\modspin)}(\eulea,\euleb,\eulec)^{\dagger}=\oprotsp^{(\modspin)}
 (\eulea,\euleb,\eulec)^{-1}=\oprotsp^{(\modspin)}(-\eulec,-\euleb,-\eulea)$
 which results from the unitarity of the rotation operators. We get:
\begin{eqnarray}\hspace{-0.15cm} \begin{array}{l}\displaystyle
 \ilkeoqe\debpscal\!\spinz\midpscal\kisoutk\!\finpscal\!=\sum_{\spinz'}
\vspace{0.1cm}\\ \displaystyle
 \ilkoeoqz\!\debvalop\!\spinz\!\midvalopa
 \oprotsp^{(\modspin)}(0,-\eulanuy,-\eulapre)\,
 \oprotsp^{(\modspin)}(-\Eularop,-\Eulanuy,-\Eulapre)
 \right.\right.\right.
\vspace{0.15cm}\\ \displaystyle \hspace{1.25cm} \times
 \left.\left.\left.\!\!
 \oprotsp^{(\modspin)}(\eulera,\eulerb,\eulerc)\,
 \oprotsp^{(\modspin)}(\Eulapre_{0},\Eulanuy_{0},\Eularop_{0})
 \midvalopb\!\spinz'\!\finvalop\irkoeoqz\;
\vspace{0.15cm}\\ \displaystyle \hspace{3.75cm} \times\,
 \ilkoeoqe\debpscal\!\spinz'\!\midpscal\!\kisin\!\finpscal\!,
 \end{array}\hspace{0.5cm}
\label{eq:rotinia} \end{eqnarray}
 where $\cZ[\vkond]=\cZ[\vkond;\Eulapre,\Eulanuy]$,
 $\cZ[\vkondini]=\cZ[\vkondini;\Eulapre_{0},\Eulanuy_{0}]$,
 $\eulea_{\entj}\equiv\eulea_{\entj}(\vkond)$ and
 $\vkond\equiv\vkond(\mkond,\eulanuy,\eulapre)$.
 The matrix element of the product of the four rotation operators
 can be calculated from the standard formula:
\begin{eqnarray} \debvalop\!\spinz\!\midvalopa\oprotsp^{(\modspin)}
 (\eulea,\euleb,\eulec)\midvalopb\!\spinz'\finvalop\!=
 \exp[-\icmp(\spinz\eulea\!+\!\spinz'\eulec)]
 \;\mrotd_{\spinz\spinz'}^{\,(\modspin)}(\euleb)\,,
 \hspace{0.75cm}
\label{eq:polpoda} \end{eqnarray}
 where $(\mrotd_{\spinz\spinz'}^{\,(\modspin)}(\euleb))$
 is a $(2\modspin+1)\!\times\!(2\modspin+1)$ matrix
 whose expression is known \cite{Mess}. 
\\

}

\paragraph{Particles of zero mass.}
{
 If the particle has zero mass, the quantization axis $\cZ[\vkond]$
 must have the same direction as that of the momentum $\vkond$.
 Since $\cz[\vkond]\parallel\vkond$ (Eq. (\!\!~\ref{eq:rotkok})),
 this implies $\cZ[\vkond]=\cz[\vkond]$.
 We then have $\Eulanuy=0$ and $\oprotos(\Eulapre,0,\Eularop)=
 \oprotos(\Eulapre+\Eularop,0,0)=\oprotos(0,0,\Eulapre+\Eularop)$ which is
 a rotation of arbitrary angle $\Eulapre+\Eularop$ around $\cz[\vkond]$.
 To simplify, we choose $\Eulapre=0$ and $\oprotos(0,0,\Eularop)$.
 We then apply the first relation of (\!\!~\ref{eq:fgenerv}) to the rotation
 $\oprotsp^{(\modspin)}(0,0,\Eularop)$.
 Using (\!\!~\ref{eq:polpoda}) and the property
 $\mrotd_{\spinz\spinz'}^{\,(\modspin)}(0)=\delta_{\spinz\spinz'}$ \cite{Mess},
 this leads to: $\debket\spinz\finket\irkeoqe=
 \exp(-\icmp\spinz\Eularop)\debket\spinz\finket\irkeoqz$.
 Then, since $\cZ[\vkond]=\cz[\vkond]$ and $\cz[\vkond]\parallel\vkond$,
 we will use the notation $\debket\spinz\finket\irkeuz$ for simplicity. Finally:
\begin{eqnarray}\debket\spinz\finket\irkeuz
 \equiv\oprotsp^{(\modspin)}(0,0,\Eularop)\debket\spinz\finket\irkeoz
 =\exp\,(-\icmp\spinz\Eularop)\debket\spinz\finket\irkeoz.\hspace{0.75cm}
\label{eq:pocirc} \end{eqnarray}
 Substituting into (\!\!~\ref{eq:rotinia}), we get:
\begin{eqnarray}\begin{array}{l}\displaystyle
 \ilkeoz\debpscal\!\spinz\midpscal\kisoutk\!\finpscal
 =\sum_{\spinz'}
\vspace{0.1cm}\\ \displaystyle \hspace{0.25cm} \times
 \,\ilkoeoz\debvalop\!\spinz\!\midvalopa
 \oprotsp^{(\modspin)}(0,-\eulanuy,-\eulapre)\,
 \oprotsp^{(\modspin)}(\eulera,\eulerb,\eulerc)
 \midvalopb\!\spinz'\!\finvalop\irkoeoz\;
\vspace{0.1cm}\\ \displaystyle \hspace{3cm} \times
 \ilkoeoz\debpscal\!\spinz'\!\midpscal\!\kisin\!\finpscal\!.
 \end{array}\hspace{0.75cm}
\label{eq:fgenero} \end{eqnarray}

 In the rest of this subsection, we apply the model to the case of the photon.
\\

}

\paragraph{Spin component amplitudes of the detected photons.}
{
 Since the photon has a spin 1, the eigenstates of its spin component
 are $\debket +1\finket$, $\debket 0\finket$, $\debket -1\finket$ but
 since it also has zero mass, the quantization axis
 is in the direction of its momentum and the eigenvalue zero is impossible
 whatever this momentum \cite{LLrel}. Therefore:
\begin{eqnarray}\ilkeoz\debpscal 0 \midpscal\kiuoutk\finpscal
 =\,\ilkoeoz\debpscal 0 \midpscal\kiuin\finpscal\,=\;0.\hspace{0.75cm}
\label{eq:fapendd} \end{eqnarray}
 This relation determines the functions
 $\eulera[\vkond(\mkond,\eulanuy,\eulapre)]$ and
 $\eulerb[\vkond(\mkond,\eulanuy,\eulapre)]$.
 Indeed, substituting it into (\!\!~\ref{eq:fgenero})
 applied to $\modspin=1$ and $\spinz=0$, we obtain:
\begin{eqnarray}\begin{array}{l}\displaystyle
 0\;=\sum_{\spinz'=\pm 1}
 \ilkoeoz\debvalop \!0\! \midvalopa
 \oprotsp^{(\spinun)}(0,-\eulanuy,-\eulapre)
 \right. \right. \right.
\vspace{0.1cm}\\ \displaystyle \hspace{1.5cm}
 \left. \left. \left. \times\,
 \oprotsp^{(\spinun)}(\eulera,\eulerb,\eulerc)
 \midvalopb\!\spinz'\!\finvalop\irkoeoz\,\ilkoeoz\debpscal\spinz'\midpscal
 \kiuin\finpscal\!,
\end{array} \hspace{0.75cm} 
\label{eq:fapende} \end{eqnarray}
 which must be satisfied whatever the initial state. Hence:
\begin{eqnarray}\ilkoeoz\debvalop\!0\!\midvalopa
 \oprotsp^{(\spinun)}(0,-\eulanuy,-\eulapre)\,
 \oprotsp^{(\spinun)}(\eulera,\eulerb,\eulerc)
 \midvalopb\!\pm 1\!\finvalop\irkoeoz\!=0.\;\hspace{0.75cm} 
\label{eq:fapendf} \end{eqnarray}
 We then express the left-hand side by using (\!\!~\ref{eq:polpoda})
 applied to $\modspin\!=\!1$ and where the matrix
 $(\mrotd_{\spinz\spinz'}^{\,(\spinun)}(\euleb))$ is given by \cite{Mess}:
%
%
\begin{eqnarray}
 \!\!\left(\mrotd_{\spinz\spinz'}^{\,(\spinun)}(\euleb)\right)\!=\!\frac{1}{2}
 \!\left( \begin{array}{ccc}
 \displaystyle 1\!+\!\cos\euleb&
 \displaystyle -\sqrt{2}\sin\euleb&
 \displaystyle 1\!-\!\cos\euleb
\vspace{0.1cm}\\
 \displaystyle \sqrt{2}\sin\euleb&
 \displaystyle 2\cos\euleb&
 \displaystyle -\sqrt{2}\sin\euleb
\vspace{0.1cm}\\
 \displaystyle 1\!-\!\cos\euleb&
 \displaystyle \sqrt{2}\sin\euleb&
 \displaystyle 1\!+\!\cos\euleb
 \end{array} \right)\!.\hspace{0.75cm}
\label{eq:djjp} \end{eqnarray}
 ({\it Note:} the order of the values of $\spinz$ and $\spinz'$ is: $+1,0,-1$).
 This leads to the equations:
\begin{eqnarray}\begin{array}{c}
 \sin\eulanuy\sin(\eulapre-\eulera)=0,
\vspace{0.2cm}\\
 \sin\eulanuy\cos\eulerb\cos(\eulapre-\eulera)-\cos\eulanuy\sin\eulerb=0.
\end{array} \hspace{0.75cm} 
\label{eq:fapendg} \end{eqnarray}
 The first equation implies $\eulera(\vkond)=\eulapre+\entn\pi$, $\entn=0,1$.
 Substituting into the second equation, we get:
 $\eulerb(\vkond)=(-1)^{\entn}\eulanuy+\entn'\pi$, $\entn'=0,1$.
 If $\eulapre=\eulanuy=0$, we then have: $\vkond=\vkondini$, which implies
 $\eulera(\vkondini)=\entn\pi$ and $\eulerb(\vkondini)=\entn'\pi$.
 But if $\vkond=\vkondini$, there is no reason for the spin polarization state
 to change. Hence,
 from (\!\!~\ref{eq:rotinid}), $\oprotsp^{(\spinun)}
 \!\left[\eulera(\vkondini),\eulerb(\vkondini),\eulerc(\vkondini)\right]$
 is equal to the identity operator, which implies:
 $\eulera(\vkondini)=\eulerb(\vkondini)=\eulerc(\vkondini)=0$.
 Therefore: $\entn=\entn'=0$ and we get:
\begin{eqnarray}\eulera(\vkond)=\eulapre,
 \hspace{0.75cm}\eulerb(\vkond)=\eulanuy, \hspace{0.75cm}
\label{eq:fapendh} \end{eqnarray}
\begin{eqnarray}\;\eulerc(\vkondini)=0. \hspace{0.75cm}
\label{eq:fapendi} \end{eqnarray}
%
 From (\!\!~\ref{eq:polpoda}), (\!\!~\ref{eq:djjp})
 and (\!\!~\ref{eq:fapendh}), the matrix whose elements appear
 in the right-hand side of (\!\!~\ref{eq:fgenero}) is given by:
\begin{eqnarray}\begin{array}{l}\displaystyle
 \!\left(\ilkoeoz\debvalop\spinz\midvalopa
 \oprotsp^{(1)}(0,-\eulanuy,-\eulapre)\,
 \oprotsp^{(1)}[\eulapre,\eulanuy,\eulerc(\vkond)]\midvalopb
 \spinz'\finvalop\irkoeoz\,\right)
\vspace{0.2cm}\\ \displaystyle \hspace{1.5cm}\;
 = \left( \begin{array}{ccc} \displaystyle
  \exp\left[-\icmp\eulerc(\vkond)\right]&\;0\;&0\vspace{0.2cm}\\ 
 0 &\;1\;& 0 \vspace{0.2cm}\\ 0 &\;0\;& \displaystyle
  \exp\left[\icmp\eulerc(\vkond)\right]
 \end{array} \,\right)\!.\hspace{0.75cm}
 \end{array}
\label{eq:chpopha} \end{eqnarray}
 Finally, from (\!\!~\ref{eq:fgenero}), (\!\!~\ref{eq:fapendh})
 and (\!\!~\ref{eq:chpopha}):
\begin{eqnarray} \ilkeoz\debpscal\!\spinz\midpscal\kiuoutk\!\finpscal\!
 =\exp\left[-\icmp\spinz\eulerc(\vkond)\,\right]\,
 \ilkoeoz\debpscal\!\spinz\midpscal\kiuin\!\finpscal. \hspace{0.75cm}
\label{eq:chpophb} \end{eqnarray}
 Diffraction causes a phase shift of $2\eulerc(\vkond)$ between
 the amplitudes of the helicity states $\debket\pm 1\finket$
 and conserves the modulus of each of these amplitudes.
\vspace{0.15cm}

}

\paragraph{Linear polarization amplitudes of the detected photons.}
{
\label{par:amplin}

 It is useful to express the amplitudes of linear polarization for any
 direction of the maximum transmission axis of the analyzer.
 We associate with the analyzer the coordinate system
 $\{\cX[\vkond],\cY[\vkond],\cz[\vkond]\}$ associated with the quantization axis
 $\cz[\vkond]$ and we assume by convention that the axis
 $\cX[\vkond]=\oprotos(0,0,\Eularop)\,\cx[\vkond]\equiv\cx[\vkond,\Eularop]$
 is the maximum transmission axis whose direction is therefore defined by
 the choice of the value of $\Eularop$.
\\

 The helicity states and the basis states of linear polarization
 in the directions of the $\ilinp[\vkond,\Eularop]$ axes ($\ilinp=\cx,\cy$)
 are related by \cite{Mess,CTDRG}:
\begin{eqnarray}\debket\helic\finket\irkeuz = \displaystyle
 \frac{-\helic}{\sqrt{2}} \left(\;\debket\xpol\finket\irkeuz
 +\icmp\,\helic\debket\ypol\finket\irkeuz\;\right), \hspace{0.75cm}
\label{eq:polinb} \end{eqnarray}
 where $\helic=\pm 1$ is the helicity.
 According to (\!\!~\ref{eq:pocirc}) applied to the helicity states
 $\debket\helic\finket\irkeuz$ and $\debket\helic\finket\irkeoz$
 expressed from (\!\!~\ref{eq:polinb}), the basis states
 $\debket\ilinp\finket\irkeoz$ transform like the real unitary vectors
 $\veclinp^{(\ilinp)}_{\vkond}$ of the $\ilinp[\vkond]$ axes:
\begin{eqnarray}\begin{array}{lll}
 \debket\xpol\finket\irkeuz
 &=&\;\,\cos\Eularop\debket\xpol\finket\irkeoz
 +\sin\Eularop\debket\ypol\finket\irkeoz,
\vspace{0.2cm}\\
 \debket\ypol\finket\irkeuz
 &=&\!\!-\sin\Eularop\debket\xpol\finket\irkeoz
 +\cos\Eularop\debket\ypol\finket\irkeoz,
\end{array}\hspace{0.75cm}
\label{eq:inelpob} \end{eqnarray}
 which implies in particular:
\begin{eqnarray}\debket\ypol\finket\irkeuz=
 \debket\xpol\finket^{}_{\!\!\vkond,\Eularop+\frac{\pi}{2}}.\hspace{0.75cm}
\label{eq:inelpof} \end{eqnarray}
 Finally, from (\!\!~\ref{eq:chpophb}), (\!\!~\ref{eq:polinb}) and
 (\!\!~\ref{eq:inelpob}), we get:
\begin{eqnarray}\begin{array}{l} \displaystyle
 \ilkeuz\debpscal\xpol\midpscal\kiuoutk\finpscal
 =\cos\left[\,\eulerc(\vkond)\!-\!\Eularop\,\right]\,
 \ilkoeoz\debpscal\!\xpol\midpscal\kiuin\finpscal
\vspace{0.2cm}\\ \displaystyle \hspace{2.5cm}\,
 \;\,-\sin\left[\,\eulerc(\vkond)\!-\!\Eularop\,\right]\,
 \ilkoeoz\debpscal\!\ypol\midpscal\kiuin\finpscal,
\end{array}\hspace{0.75cm}
\label{eq:polinc} \end{eqnarray}
 from which we deduce $\,\ilkeuz\debpscal\ypol\midpscal\kiuoutk\finpscal$
 by using (\!\!~\ref{eq:inelpof}).
\\

}

\paragraph{Case of an initial state elliptically polarized (photons).}
{
 By generalizing (\!\!~\ref{eq:polinb}), we can express
 any elliptically polarized initial state in the form:
\begin{eqnarray}
 \debket\etaelli^{(\spinun)}_{\textup{in}}\!\finket\!\!
 \equiv -\helicin\!\left(\,\cos\ellipin \debket\xpol\finket\irkoeiz
 \!+\icmp\,\helicin\sin\ellipin \debket\ypol\finket\irkoeiz\,\right)\!,
 \hspace{0.75cm}
\label{eq:inelpoa} \end{eqnarray}
 where $\azaxeli$, $\ellipin$ and $\helicin$ represent respectively
 the major axis azimuth, the ellipticity angle and the handedness
\footnote{
 We use the following definitions:
 
 $\azaxeli\equiv\azaxel(\vkondini)=$ angle between the $\cx[\vkondini]$ axis
 and the major axis of the ellipse in the transverse plane to $\vkondini$,
 $0\leq\azaxeli<\pi$.
 
 $\ellipin=\arctan$[\,(length of the minor axis)/(length of the major axis)\,],
 $0\leq\ellipin\leq\pi/4$.
 
 $\helicin=\pm 1$, represents the direction of rotation of the
 electric field vector (provided that $\ellipin\neq 0$).
 The value $\helicin=+1$ corresponds to a counterclockwise
 rotation if the rotation axis and the momentum
 of the photon are directed toward the receiver.
 
 If $\ellipin=0$, the polarization is linear along the direction defined by
 the angle $\azaxeli$.
 If $\ellipin=\pi/4$, the polarization is circular and $\helicin$ is equal
 to the helicity because (\!\!~\ref{eq:inelpoa}) becomes identical to
 (\!\!~\ref{eq:polinb}) applied to $\helic=\helicin$,
 $\vkond=\vkondini$ and $\Eularop=\azaxeli$.
%
}.

 The final state resulting from the initial state
 $\debket\etaelli^{(\spinun)}_{\textup{in}}\finket$
 is also an elliptically polarized state which we denote
 $\debket\etaelli^{(\spinun)}_{\textup{out}}(\vkond)\finket$.
 Indeed, by applying (\!\!~\ref{eq:polinc}) to
 $\debket\etaelli^{(\spinun)}_{\textup{in}}\finket$ defined by
 (\!\!~\ref{eq:inelpoa}) and using (\!\!~\ref{eq:inelpob}), we obtain:
\begin{eqnarray}\begin{array}{rcl} \displaystyle
 \ilkeuz\debpscal\xpol\midpscal
 \etaelli^{(\spinun)}_{\textup{out}}(\vkond)\finpscal
 &\!\!=\!\!&-\helicin\cos\ellipin
 \cos\left[\,\azaxeli\!+\!\eulerc(\vkond)\!-\!\Eularop\,\right]\;
\vspace{0.1cm}\\
 &&\;\;+\;\icmp\,\sin\ellipin
 \sin\left[\,\azaxeli\!+\!\eulerc(\vkond)\!-\!\Eularop\,\right].
\end{array} \hspace{0.75cm}
\label{eq:inelpoc} \end{eqnarray}
 Then, by making the identity operator
 $\sum_{\ilinp=\cx,\cy}\debket\ilinp\finket\irkeoz\,
 \ilkeoz\debbra\ilinp\finbra$ act on the state
 $\debket\etaelli^{(\spinun)}_{\textup{out}}(\vkond)\finket$
 and using successively (\!\!~\ref{eq:inelpoc})
 (applied with $\Eularop=0$), (\!\!~\ref{eq:inelpof})
 and (\!\!~\ref{eq:inelpob}), we get:
\begin{eqnarray}\begin{array}{l}\displaystyle
 \debket\etaelli^{(\spinun)}_{\textup{out}}(\vkond)\finket
 =-\helicin\left[\, \cos\ellipin
 \debket\xpol\finket^{}_{\!\!\vkond,\azaxeli+\eulerc(\vkond)}
\right.\vspace{0.1cm}\\ \left.\displaystyle \hspace{3cm}
 +\,\icmp\,\helicin\,\sin\ellipin
 \debket\ypol\finket^{}_{\!\!\vkond,\azaxeli+\eulerc(\vkond)}
 \,\right]. \end{array} \hspace{0.75cm}
\label{eq:inelpog} \end{eqnarray}
 Comparing with (\!\!~\ref{eq:inelpoa}), we see that the ellipticity
 and the handedness are conserved and that the ellipse axes undergo
 a rotation of angle $\eulerc(\vkond)$.
 The major axis azimuth in the transverse plane $\{\cx[\vkond],\cy[\vkond]\}$
 is: $\azaxel(\vkond)=\azaxeli+\eulerc(\vkond)$.

}

}

}

\section{Some predictions of the model}
{
\label{sec:predreli}

\subsection{Relative intensity (polarization not measured)}
{
\label{sub:densprob}

\paragraph{Angular distribution of the final momentum.}
{
 From  (\!\!~\ref{eq:fgenerq}) and (\!\!~\ref{eq:opdispa}), the p.d.f. of
 the final momentum if the polarization is not measured is expressed by:
\begin{eqnarray} \begin{array}{l}\displaystyle
 \pdfp^{}_{\vavkond}(\vkond)\,\simeq\,\norm^{-1}\,(2\pi)^{-4}\mkondini^{-2}
\vspace{0.2cm}\\ \displaystyle \hspace{1.5cm} \times\,
 \gaussmk\!\left(|\vkond|\!-\!\mkondini\right)\delta_{1\,\sgn[\kz]}
 \left|\jfour^{\ouva}(\vkond\!-\!\vkondini)\right|^2\!\!.
\end{array} \hspace{0.75cm}
\label{eq:dpkcart} \end{eqnarray}

 Since the experimental setup directly measures the direction of $\vkond$,
 it is useful to replace the Cartesian components
 by the modulus and two angles giving the direction.
 This change of variables must be done by a one-to-one transformation which must
 moreover be defined in the half-space $\kz>0$ because of (\!\!~\ref{eq:contray}).
 The spherical coordinates $\mkond,\angdev,\eulapre$ cannot be used because
 the associated transformation is not one-to-one (if $\angdev=0$,
 $\eulapre$ is undetermined and the Jacobian is zero).
 On the other hand, we can use the {\it diffraction angles} $\angdif_{\cx}$
 and $\angdif_{\cy}$ \cite{LLangdif} which are the projections
 of the polar angle $\angdev$ on the planes
 $(\cx,\cz)$ and $(\cy,\cz)$ (Fig. \!\!~\ref{fig:dirang}).
%
\begin{figure}[h]\hspace{-0.75cm}
\begin{minipage}[t]{8.75cm}
 \resizebox{1.1\textwidth}{!}{%
\includegraphics{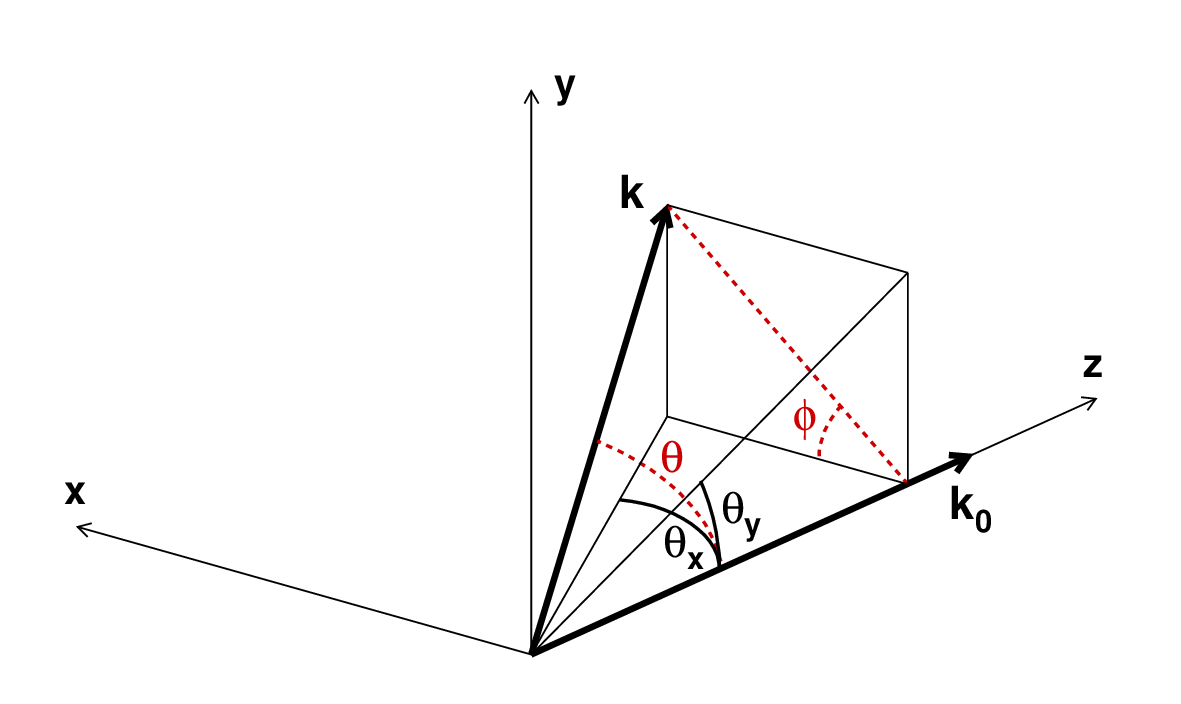}}
\vspace{-0.5cm}
 \caption{
 Diffraction angles $\angdif_{\cx}$ and $\angdif_{\cy}$.}
 \label{fig:dirang}
 \end{minipage}
\end{figure}
%
 The new variables
 $(\mkond,\angdif_{\cx},\angdif_{\cy})$ are such that: $\mkond>0$,
 $-\pi/2<\angdif_{\cx}<+\pi/2$, $-\pi/2<\angdif_{\cy}<+\pi/2$ and
 the required transformation $(\kx,\ky,\kz)\leftrightarrow(\mkond,\angdif_{\cx},
 \angdif_{\cy})$ is:
\begin{eqnarray}\begin{array}{c}
 \hspace{0.75cm}\vkond\left(\mkond,\angdif_{\cx},\angdif_{\cy}\right)
 \;=\;\mkond\,\cos\angdev\,
 \left(\begin{array}{c}
 \tg\angdif_{\cx}\vspace{0.075cm} \\ \tg\angdif_{\cy} \vspace{0.075cm}\\ 1
\end{array}\right),
\vspace{0.2cm} \\
 \cos\angdev=\left(1\!+\!\tg^2\angdif_{\cx}\!+\!\tg^2\angdif_{\cy}
 \right)^{-1/2}\!,\hspace{0.25cm} 0\leq\angdev<\pi/2.
 \end{array}\hspace{0.75cm} 
\label{eq:ouvrb} \end{eqnarray}

 The change of p.d.f. due to the change of variables is expressed by:
\begin{eqnarray}\!\pdfp^{}_{\vamkond,\vadifx,\vadify}
 (\mkond,\angdif_{\cx},\angdif_{\cy})
 =\left|\jacobien(\mkond,\angdif_{\cx},\angdif_{\cy})\right|
 \pdfp^{}_{\vavkond}[\,\vkond(\mkond,\angdif_{\cx},\angdif_{\cy})\,],
 \hspace{0.75cm}
\label{eq:ouvrf} \end{eqnarray}
 where $\jacobien(\mkond,\angdif_{\cx},\angdif_{\cy})$ is the determinant
 of the Jacobian of the transformation (\!\!~\ref{eq:ouvrb}) which is finite
 and non-zero and whose calculation leads to the {\it angular factor:}
\begin{eqnarray}\fgeomang\left(\angdif_{\cx},\angdif_{\cy}\right)
 \equiv\mkond^{-2}\left|\jacobien(\mkond,\angdif_{\cx},\angdif_{\cy})\right|=
 \frac{\cos\angdev}{1-\sin^2\!\angdif_{\cx}\,\sin^2\!\angdif_{\cy}}.
 \hspace{0.75cm}
\label{eq:fganga} \end{eqnarray}
 Expressing $\pdfp^{}_{\vavkond}\left[\,
 \vkond(\mkond,\angdif_{\cx},\angdif_{\cy})\,\right]$
 from (\!\!~\ref{eq:dpkcart}) and substituting
 into (\!\!~\ref{eq:ouvrf}), given (\!\!~\ref{eq:fganga}), we get:
\begin{eqnarray}\!\!\begin{array}{l}\displaystyle
 \pdfp^{}_{\vamkond,\vadifx,\vadify}\!\left(\mkond,\angdif_{\cx},
 \angdif_{\cy}\right)\simeq\norm^{-1}(2\pi)^{-4}\mkondini^{-2}
 \mkond^2\,\gaussmk\!\left(\mkond\!-\!\mkondini\right)
\vspace{0.2cm} \\ \displaystyle \hspace{1.25cm}\times\,
 \fgeomang(\angdif_{\cx},\angdif_{\cy})\left|\,\jfour^{\ouva}
 \!\left[\,\vkond(\mkond,\angdif_{\cx},\angdif_{\cy})\!-\!
 \vkond(\mkondini,0,0)\,\right]\,\right|^2\!\!.
 \end{array}\hspace{0.75cm}
\label{eq:mcfoh} \end{eqnarray}
 From (\!\!~\ref{eq:contrax}), $\Delta\mkond$ is close to zero.
 We can therefore replace the function $\gaussmk\left(\mkond-\mkondini\right)$
 by the Dirac distribution $\delta(\mkond-\mkondini)$
 and express the angular distribution of the final momentum by: 
\begin{eqnarray}\!\!\begin{array}{l}\displaystyle
 \pdfp^{}_{\vadifx,\vadify}\!\left(\angdif_{\cx},\angdif_{\cy}\right)
 \equiv\!\int_{0}^{\infty}\!\!\!\differ\mkond'\,
 \pdfp^{}_{\vamkond,\vadifx,\vadify}\!\left(\mkond',\angdif_{\cx},
 \angdif_{\cy}\right)
\\ \\ \displaystyle \hspace{0.75cm}
 \simeq\,\frac{\fgeomang(\angdif_{\cx},\angdif_{\cy})}{(2\pi)^{4}\norm}\,
 \left|\,\jfour^{\ouva}\!\left[\,\vkond(\mkond,\angdif_{\cx},\angdif_{\cy})
 -\vkond(\mkond,0,0)\,\right]\,\right|^2.
\end{array}\hspace{0.75cm}
\label{eq:mcfou} \end{eqnarray}
 where we now consider for simplicity that $\mkond$ represents both
 the modulus of $\vkondini$ and that of $\vkond$.
 The normalization factor $\norm$ can be expressed by substituting
 (\!\!~\ref{eq:opdispa}) into (\!\!~\ref{eq:fgenerw}).
 Using the change of variables (\!\!~\ref{eq:ouvrb})
 and given (\!\!~\ref{eq:contrax}), we get:
\begin{eqnarray}\begin{array}{l}\displaystyle
 \norm\simeq(2\pi)^{-4}\!\!
 \int_{-\pi/2}^{+\pi/2}\!\!\differ\angdif_{\cx}
 \int_{-\pi/2}^{+\pi/2}\!\!\differ\angdif_{\cy}
 \vspace{0.2cm}\\ \displaystyle \hspace{1cm} \times\,
 \fgeomang(\angdif_{\cx},\angdif_{\cy})\left|\,\jfour^{\ouva}
 \!\left[\,\vkond(\mkond,\angdif_{\cx},\angdif_{\cy})
 -\vkond(\mkond,0,0)\,\right]\,\right|^2\!\!\!.
 \end{array}\hspace{0.75cm}
\label{eq:mcfon} \end{eqnarray}

}

\paragraph{Quantum formula of the relative intensity in Fraunhofer scalar diffraction.}
{
 To avoid calculating the integral (\!\!~\ref{eq:mcfon}), we consider
 the ratio of the values of the angular distribution between the direction
 $(\angdif_{\cx},\angdif_{\cy})$ and the forward direction $(0,0)$.
 This ratio is nothing other than the relative intensity between
 the directions of $\vkond$ and $\vkondini$.
 Thus, in the quantum model (QM), the expression of the the relative intensity is:
\begin{eqnarray} \left[\,\frac{\intens(\angdif_{\cx},\angdif_{\cy})}
 {\intens(0,0)}\,\right]^{\ouva}_{\textup{QM}}\;=\;
 \frac{\pdfp^{}_{\vadifx,\vadify}(\angdif_{\cx},\angdif_{\cy})}
      {\pdfp^{}_{\vadifx,\vadify}(0,0)}.
\label{eq:intrela} \end{eqnarray}
 From (\!\!~\ref{eq:mcfou}) and since $\fgeomang(0,0)=1$, this leads to:
\begin{eqnarray}\begin{array}{l}\displaystyle
\!\!\left[\,\frac{\intens(\angdif_{\cx},\angdif_{\cy})}
 {\intens(0,0)}\,\right]^{\ouva}_{\mathrm{QM}}
 \simeq\;\fgeomang(\angdif_{\cx},\angdif_{\cy})
\\ \displaystyle \hspace{2cm}\times\;
 \frac{\displaystyle\left|\,
 \jfour^{\ouva}[\,\vkond(\mkond,\angdif_{\cx},\angdif_{\cy})\!-\!
 \vkond(\mkond,0,0)\,]
 \,\right|^2}{\displaystyle \left|\,\jfour^{\ouva}\left(0\right)\right|^2}.
\end{array}\hspace{0.75cm}
\label{eq:omonoch} \end{eqnarray}
 For an aperture of the form $\ouva\equiv\ouvas\times[-\Delta\cz/2,+\Delta\cz/2]$,
 where $\Delta\cz$ is independent of $(\cx,\cy)$, the position filtering function
 $\fredA(\rvec)$ is equal to $\fred^{\ouvas,\Delta\cz}(\rvec)$ given by
 (\!\!~\ref{eq:tronqfoc}). From this and (\!\!~\ref{eq:tronqfod}), the relation
 (\!\!~\ref{eq:nrspinb}) leads to:
\begin{eqnarray}
 \jfour^{\ouva}(\vkond-\vkondini)\;=\;
 \jfour^{\ouvas}_{\indtran}(\kx,\ky)\;\jfour^{\Delta\cz}_{\indlong}(\kz-\mkond),
\label{eq:prpobf} \end{eqnarray}
 where:
\begin{eqnarray}\begin{array}{l} \displaystyle
 \jfour^{\ouvas}_{\indtran}(\kx,\ky)\,\equiv\,(2\pi)^{-1}\,\surfouva^{-1/2}
\vspace{0.2cm}\\ \displaystyle \hspace{2.25cm}\times
 \!\!\intdbl_{\ouvas}\!\!\differ\cx\differ\cy\,\;
 \exp\left[\,-\icmp(\kx\cx+\ky\cy)\,\right],
\end{array}\hspace{0.75cm}
\label{eq:prpobc} \end{eqnarray}
\begin{eqnarray}\begin{array}{l} \displaystyle
 \jfour^{\Delta\cz}_{\indlong}(\kz-\mkond)\,\equiv\, (2\pi)^{-1/2}
\vspace{0.2cm}\\ \displaystyle \hspace{2cm}\times
 \!\!\int\!\differ\cz\;
 \sqrt{\fredDzl(\cz)} \;\exp\left[\,-\icmp(\kz-\mkond)\cz\,\right].
\end{array}\hspace{0.75cm}
\label{eq:prpobd} \end{eqnarray}
 Substituting (\!\!~\ref{eq:prpobf}) into (\!\!~\ref{eq:omonoch})
 and expressing $\vkond(\mkond,\angdif_{\cx},\angdif_{\cy})$ from
 ($\!\!$~\ref{eq:ouvrb}), we obtain:
\begin{eqnarray}
 \hspace{-0.25cm}\left[\frac{\intens(\angdif_{\cx},\angdif_{\cy})}
 {\intens(0,0)}\right]^{\ouva}_{\mathrm{QM}}
 \!\!\simeq\fgeomang(\angdif_{\cx},\angdif_{\cy})\,
 \tdiftran^{\ouvas}(\mkond,\angdif_{\cx},\angdif_{\cy})\,
 \tdiflong^{\Delta\cz}(\mkond,\angdev),\hspace{0.5cm}
\label{eq:intrelm} \end{eqnarray}
 where $\angdev$ and $\fgeomang(\angdif_{\cx},\angdif_{\cy})$
 are respectively given by (\!\!~\ref{eq:ouvrb}) and (\!\!~\ref{eq:fganga}),
 $\tdiftran^{\ouva}(\mkond,\angdif_{\cx},\angdif_{\cy})$
 is the {\it transverse diffraction term}:
\begin{eqnarray}\hspace{-0.5cm}\begin{array}{l}
 \tdiftran^{\ouvas}(\mkond,\angdif_{\cx},\angdif_{\cy})
 \equiv\frac{\displaystyle\,\left|\,
       \jfour^{\ouvas}_{\indtran}(\mkond\cos\angdev\,\tg\angdif_{\cx}\,,\,
       \mkond\cos\angdev\,\tg\angdif_{\cy})\,\right|^2}
      {\displaystyle\left|\,\jfour^{\ouvas}_{\indtran}(0,0)\,\right|^2}
\end{array}\hspace{0.25cm}
\label{eq:intrelt} \end{eqnarray}
 and $\tdiflong^{\ouva}(\mkond,\angdev)$
 is the {\it longitudinal diffraction term}:
\begin{eqnarray}\tdiflong^{\Delta\cz}(\mkond,\angdev)
 \;\equiv\;\frac{\displaystyle
 \left|\,\jfour^{\Delta\cz}_{\indlong}[\,\mkond\,(\cos\angdev-1)\,]\,\right|^2}
      {\displaystyle\left|\,\jfour^{\Delta\cz}_{\indlong}(0)\,\right|^2}.
\label{eq:intreln} \end{eqnarray}
%

}

\paragraph{Test of the Huygens-Fresnel principle.}
{
 The relative intensity expressed by the quantum formula
 (\!\!~\ref{eq:intrelm}) depends on the width $\Delta\cz$
 of the longitudinal 1D aperture (Fig. \!\!~\ref{fig:ouvgenz}).
 The value of $\Delta\cz$ can therefore be fitted to data obtained
 from the measurement of the intensity as a function of the diffraction angle.
 As previously mentioned (\S \!\!~\ref{par:posfilt}),
 $\Delta\cz$ is the width of the distribution of the wavefronts
 emitting the wavelets which contribute to the diffracted wave.
 An experimental study directly concerning the Huygens-Fresnel principle
 can therefore be considered.
\\

}

\paragraph{Comparison with the predictions of the scalar theories of wave optics.}
{
 In wave optics (WO), there are several versions of the scalar theory of
 diffraction which differ by their assumed boundary conditions.
 The best known are the theories of
 Fresnel-Kirchhoff (FK) and Rayleigh-Sommerfeld (RS1 and RS2).
 In Fraunhofer diffraction, for an initial monochromatic plane wave
 in normal incidence,
 the amplitude predicted by these theories at a point of radius vector
 $\vdistdipo$ beyond the diaphragm can be expressed,
 given (\!\!~\ref{eq:detadn}), in the form \cite{BoWo,Soma}:
\begin{eqnarray}\begin{array}{l}\displaystyle
 \amplik^{\ouvas}(\vdistdipo)\equiv
 \amplik^{\ouvas,\mkond}\!\left(\distdipo,\frac{\vkond}{\mkond}\right)\!
 \simeq-\,\vargc_{0}\frac{\icmp\mkond}{2\pi}
 \frac{\exp\left[\icmp\mkond
 \!\left(\distsodi\!+\!\distdipo\right)\right]}{\distsodi\,\distdipo}
\vspace{0.2cm}\\ \displaystyle \hspace{0.5cm}
 \times\,\facincl[(\vkondini,\vkond)]
 \!\intdbl_{\ouvas}\differ\cx\differ\cy\;
 \exp\!\left[-\icmp\mkond\left(\frac{\kx}{\mkond}\cx
 +\frac{\ky}{\mkond}\cy\right)\right],
 \end{array} \hspace{0.75cm}
\label{eq:cfrkirl} \end{eqnarray}
 where $\vargc_{0}$ is a constant, $\distsodi$ is the distance source-aperture
 and $\facincl[(\vkondini,\vkond)]$ is the {\it obliquity factor}.
 The latter depends on the {\it deflection angle} $(\vkondini,\vkond)$
 which is also the polar angle $\angdev$ (Fig. \!\!~\ref{fig:dirang}).
 Its value is specific to the theory:
\begin{eqnarray}\facincl(\angdev)\;=\;
 \left\{ \begin{array}{cc}\displaystyle
 (1+\cos\angdev)/2 \hspace{1cm}&\mbox{(FK)}
\vspace{0.05cm}\\ \displaystyle\cos\angdev \hspace{1cm}&\mbox{(RS1)}
\vspace{0.05cm}\\ \displaystyle 1 \hspace{1cm}&\;\mbox{(RS2)}.
\end{array}\right.\hspace{0.75cm}
\label{eq:cfrkirfi} \end{eqnarray}

 From (\!\!~\ref{eq:detadn}), the intensity at point of radius-vector
 $\vdistdipo$ is proportional to the intensity in the direction of
 $\vkond(\mkond,\angdif_{\cx},\angdif_{\cy})$. Hence:
\begin{eqnarray}\left[\,\frac{\intens(\angdif_{\cx},\angdif_{\cy})}
 {\intens(0,0)}\,\right]^{\ouvas}_{\mathrm{WO}}=\;
 \frac{\left|\,\amplik^{\ouvas,\mkond}
 \left(\distdipo,\frac{\vkond(\mkond,\angdif_{\cx},\angdif_{\cy})}
 {\mkond}\right)\,\right|^2}{\left|\,\amplik^{\ouvas,\mkond}
 \left(\distdipo,\frac{\vkond(\mkond,0,0)}{\mkond}\right)\,\right|^2}.
 \hspace{0.75cm}
\label{eq:intrelka} \end{eqnarray}
 Expressing $\vkond(\mkond,\angdif_{\cx},\angdif_{\cy})$
 and $\vkond(\mkond,0,0)$ from ($\!\!$~\ref{eq:ouvrb}) and
 substituting into ($\!\!$~\ref{eq:cfrkirl}) then into
 ($\!\!$~\ref{eq:intrelka}), we see that $\distdipo$ is eliminated.
 Then, since $\facincl(0)=1$ and given ($\!\!$~\ref{eq:prpobc}) and
 ($\!\!$~\ref{eq:intrelt}):
\begin{eqnarray}\left[
 \frac{\intens(\angdif_{\cx},\angdif_{\cy})}
 {\intens(0,0)}\right]^{\ouvas}_{\mathrm{WO}}
 \simeq\;\facincl(\angdev)^2\;\,
 \tdiftran^{\ouvas}(\mkond,\angdif_{\cx},\angdif_{\cy}).
\label{eq:intrell} \end{eqnarray}

 The comparison of formulae (\!\!~\ref{eq:intrelm}) and (\!\!~\ref{eq:intrell})
 shows that the transverse diffraction term
 $\tdiftran^{\ouvas}(\mkond,\angdif_{\cx},\angdif_{\cy})$ is the same
 in the two cases.
 This is because the integrals in (\!\!~\ref{eq:prpobc}) and
 (\!\!~\ref{eq:cfrkirl}) are the same.
 The differences come from the angular factors
 $\fgeomang(\angdif_{\cx},\angdif_{\cy})$ and $\facincl(\angdev)^2$ and from
 the presence of the longitudinal diffraction term
 $\tdiflong^{\Delta\cz}(\mkond,\angdev)$ in the quantum formula.
 If the angles are small, the angular factors and the
 longitudinal diffraction term are all close to 1 so that the quantum model
 gives the same result as that of wave optics.
 On the other hand, if the angles increase,
 discrepancies appear between the different predictions.
\\

}

\paragraph{Example of comparison.}
{
 Let us consider the intensity variation in the horizontal plane $(Ox,Oz)$
 for which we have: $\angdif_{\cy}=0$, $\angdif_{\cx}=\angdev$
 if $\angdif_{\cx}\geq 0$, $\angdif_{\cx}=-\angdev$ if $\angdif_{\cx}\leq 0$.
 In this case, it is convenient to make the notation change:
 $(\angdif_{\cx},\angdev)\rightarrow(\angdif,|\angdev|)$, where
 $-\pi/2<\angdif<+\pi/2$ (diffraction angle) and
 $0\leq|\angdev|<+\pi/2$ (polar angle in the half-space $\cz>0$).
 Since $\cos|\angdev|=\cos\angdev$, the relations
 (\!\!~\ref{eq:fganga}) and (\!\!~\ref{eq:cfrkirfi}) then lead to:
\begin{eqnarray}\fgeomang(\angdif,0)=\cos\angdev,\hspace{0.75cm}
 \facincl(|\angdev|)=\facincl(\angdev).\hspace{0.75cm}
\label{eq:difrect} \end{eqnarray}

 We now consider the case of a rectangular slit $\fente$ of width $2\larg$
 and of height $2\haut$ centered at $(\cx,\cy)=(0,0)$.
 The expression (\!\!~\ref{eq:prpobc}) leads to:
\begin{eqnarray}\jfour^{\fente}_{\indtran}(\kx,\ky)\,=\,
 \frac{\sqrt{\larg\haut}}{\pi}\;\frac{\sin\larg\kx}{\larg\kx}
 \;\frac{\sin\haut\ky}{\haut\ky}. \hspace{0.75cm}
\label{eq:fourtr} \end{eqnarray}
 Given the notation change introduced above, the relation (\!\!~\ref{eq:ouvrb})
 implies: $\kx=\mkond\cos|\angdev|\tg\angdif=\mkond\sin\angdev$ and $\ky=0$.
 Applying (\!\!~\ref{eq:fourtr}) to these values
 and substituting into (\!\!~\ref{eq:intrelt}), we get the well-known result:
\begin{eqnarray}\tdiftran^{\fente}(\mkond,\angdif,0)\;=\;
 \left[\,\frac{\sin(\larg\mkond\sin\angdif)}
 {\larg\mkond\sin\angdif}\,\right]^2.\hspace{0.75cm}
\label{eq:ttransr} \end{eqnarray}

 Then, we suppose that the longitudinal filtering function is for example
 a Gaussian. In this case, the width of the longitudinal aperture depends
 on the standard deviation and on a threshold under which the integral of the
 Gaussian outside the interval $[-\Delta\cz(\sigz)/2,+\Delta\cz(\sigz)/2]$
 is considered as negligible (for example, with a threshold of $10^{-2}$,
 we have: $\Delta\cz(\sigz)\simeq 5.16\,\sigz$ \cite{pdgGauss}).
 Assuming that $\fred_{\indlong}^{\Delta\cz(\sigz)}(\cz)$
 is a Gaussian centered at $\cz=0$ and of standard deviation $\sigz$,
 the expression (\!\!~\ref{eq:prpobd}) leads to \cite{GradRyz}:
\begin{eqnarray}
 \jfour^{\Delta\cz(\sigz)}_{\indlong}(\kz\!-\!\mkond)=
 \!\left(\frac{2}{\pi}\right)^{\!\!1/4}\!\!\!\!\sqrt{\sigz}\,
 \exp\!\left[-{\sigz}^2(\kz\!-\!\mkond)^2\right]\!.\hspace{0.75cm}
\label{eq:poiga} \end{eqnarray}
 Substituting into (\!\!~\ref{eq:intreln}), we get:
\begin{equation}\tdiflong^{\Delta\cz(\sigz)}(\mkond,|\angdev|)\;=\;
 \exp\left[\,-8\,{\sigz}^2\mkond^2\,\sin^4(\angdif/2)\,\right].
\label{eq:tlongig} \end{equation}
\vspace{0.1cm}

 Curves obtained from formulae (\!\!~\ref{eq:intrelm})
 and (\!\!~\ref{eq:intrell}) (applied with
 (\!\!~\ref{eq:cfrkirfi}), (\!\!~\ref{eq:difrect}),
 (\!\!~\ref{eq:ttransr}) and (\!\!~\ref{eq:tlongig}))
 are shown in Fig. \!\!\!\!~\ref{fig:diffraca}
 for a case of photon diffraction.
\\

 If $\sigz=0$, the longitudinal diffraction term is equal to 1.
 This corresponds to the largest values predicted by the quantum model.
 It is with the FK theory that the quantum model (QM1)
 is in better agreement.
 However, at $90^{\circ}$, the FK theory predicts
 values that are generally non-zero,
 which does not seem plausible (same for the RS2 theory).
 The angular factors $\fgeomang(\angdif,0)=\cos\angdev$ of the quantum model
 and $\facincl(\angdev)^2=\cos^{2}\angdif$ of the RS1 theory 
 are the only ones which account for
 the decrease in intensity towards zero at 90$^{\circ}$.
 However, the factor $\cos\angdif$ seems more likely because
 it is the same as that obtained by applying the exact calculation
 of the diffraction by a wedge \cite{Somb} to the case of two wedges
 of zero angle placed opposite one another to form a slit \cite{LLSom}.
\\

 If $\sigz>0$, the longitudinal diffraction term is strictly less than 1.
 The values of the quantum model, maximum for $\sigz=0$, undergo a damping
 which increases with $|\angdif|$ and $\sigz$.
 As $\sigz$ increases from zero, the QM curve deviates more and more
 from the QM1 curve and then goes below the RS1 curve.
 Coincidentally, the curves QM and RS1 can be very close
 but not for all values of $\angdif$ since the angular factors are different.
 If $\sigz$ is large enough, the QM curve globally decreases much more rapidly
 than the WO and QM1 curves and the gap becomes significant
 at not too large angles (QM2).
 Such a result obtained experimentally would be a signal of the need
 to use a "multi-wavefronts" Huygens-Fresnel principle
 to describe the diffraction by an aperture.
\\

\begin{figure}[t]
\begin{minipage}{8.5cm}
\centering\hspace{-1.cm}
\resizebox{1.1\textwidth}{!}{
 \includegraphics[width=1.26\textwidth, height=1.55\textwidth]
 {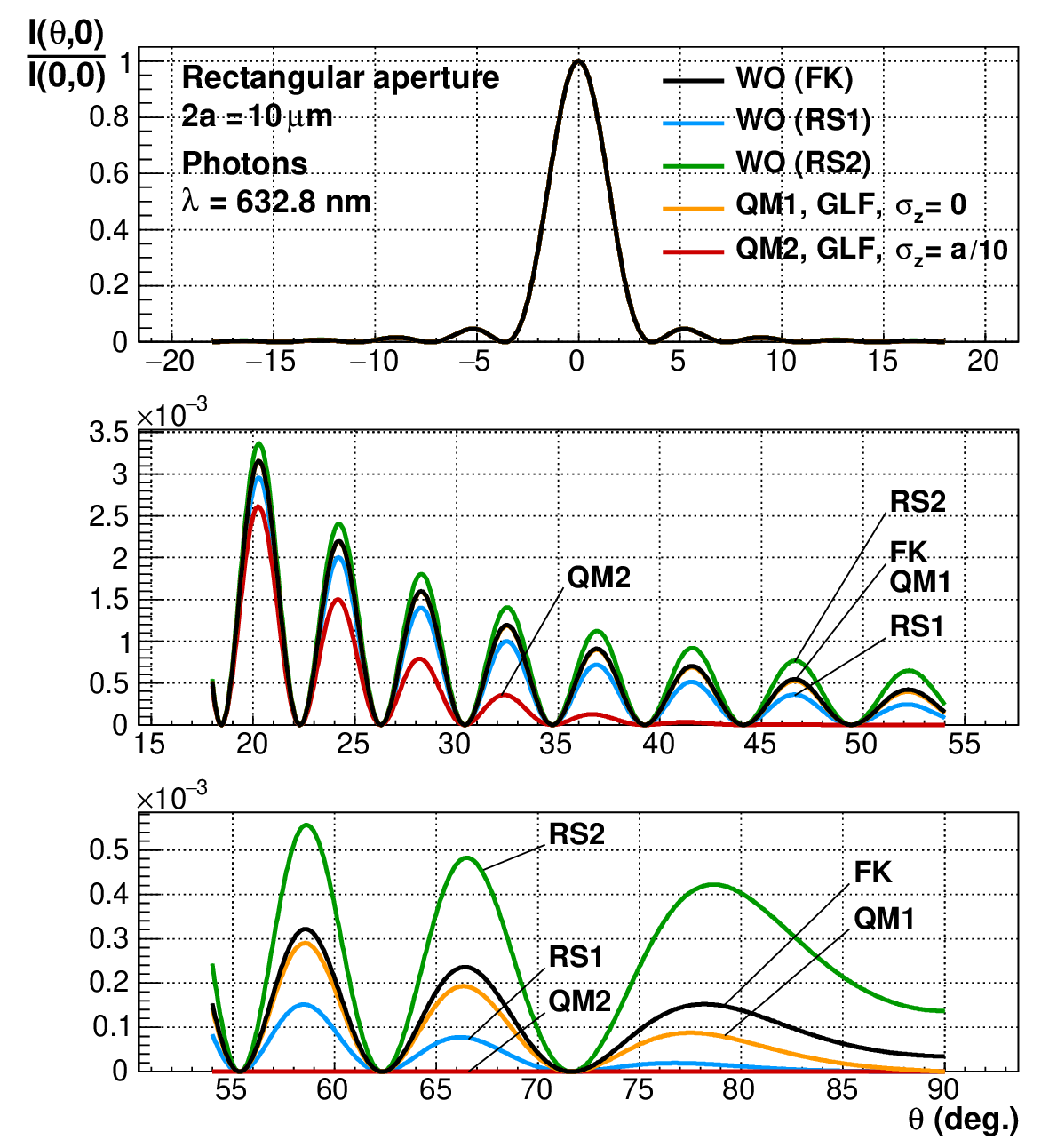}
}
\end{minipage}\hfill
\begin{minipage}{8.5cm}
 \caption{
 Comparison between different theoretical predictions 
 of the relative intensity in Fraunhofer diffraction
 as a function of the diffraction angle in the horizontal plane
 for a rectangular slit of width $2\larg=10\;\mu$m and
 an incident monochromatic plane wave corresponding to photons
 of wavelength $\lgond=632.8$ nm (Helium-Neon laser).
 Five predictions are presented: three predictions of wave optics (WO)
 corresponding to the scalar theories of Fresnel-Kirchhoff (FK) and
 Rayleigh-Sommerfeld (RS1 and RS2) and two predictions of the quantum model
 (QM) corresponding to two values of the standard deviation $\sigz$
 associated with a Gaussian longitudinal filtering (GLF) of the incident wave:
 $\sigz=0$ (QM1) and $\sigz=\larg/10$ (QM2).
 The values of the five intensities are distributed according to the decreasing
 order: RS2, FK, QM1, RS1, QM2 over the whole range 0$^\circ$- 90$^\circ$.
 These predictions correspond to the case where the polarization is not measured.
}
\label{fig:diffraca}
\end{minipage}
\end{figure}

}

\paragraph{Large diffraction angles.}
{
 From the above analysis, it turns out
 that the relative gaps between the predictions of the different models
 considered here are significant at large angles.
 Moreover, from a survey of the literature, it seems that
 no accurate experimental study of the diffraction in this region
 has been carried out so far.
 Since the time when the FK and RS1-2 theories were formulated
 (late nineteenth century),
 technologies in optics have made tremendous progress
 thanks in particular
 to accurate measurements of intensity by charge-coupled
 devices which make it possible to achieve
 a sufficiently expanded dynamic range.
 An experimental study of this still little explored region
 is therefore probably feasible at the present time.

}

}

\subsection{Polarization probabilities (photons)}
{
\label{sub:propolar}

 From (\!\!~\ref{eq:fgenerr}) and (\!\!~\ref{eq:chpophb}), the conditional
 probability to detect a photon of helicity $\helic$ if its momentum is
 $\hbar\vkond$ is:
\begin{eqnarray}
 \!\!\probcon_{\vaspinzh}^{(\spinun)}\!\left([\helic]_{\vkond}\right)
 =\left|\!\ilkeoz\debpscal\!\helic\midpscal
 \kiuoutk\!\finpscal\!\right|^2\!\!
 =\left|\!\ilkoeoz\debpscal\!\helic\midpscal\!\kiuin
 \finpscal\!\right|^2\!\!.
 \hspace{0.75cm}
\label{eq:chpophc} \end{eqnarray}
 So the probabilities of the helicity states and consequently of the circular
 polarizations are conserved.

 {\it Note:}
 for an aperture of sub-wavelength size, circular polarization probabilities
 are not conserved for all diffraction angles because the aperture limits the
 transmission of circularly polarized light \cite{Shin}.
 This effect is not taken into account in assumption (\!\!~\ref{eq:rotinid})
 and consequently the polarization predicted by the model does not match to
 experiment in this specific case.
\\

 For an elliptically polarized initial state
 $\debket\etaelli^{(\spinun)}_{\textup{in}}\finket$,
 with major axis azimuth $\azaxeli$, ellipticity angle $\ellipin$ and
 handedness $\helicin$ (Eq. (\!\!~\ref{eq:inelpoa})),
 the conditional probabilities of linear polarization
 in the direction defined by the angle $\Eularop$
 with respect to the $\cx[\vkond]$ axis
 are expressed, from (\!\!~\ref{eq:inelpoc}), by:
\begin{eqnarray}\!\!\begin{array}{l}\displaystyle
 \probcon_{\vaxpxizh}^{(\spinun)}\left([\xpol]_{\vkond,\Eularop}\right)
 =\left|\;\ilkeuz\debpscal\xpol\midpscal
 \etaelli^{(\spinun)}_{\textup{out}}(\vkond)\finpscal\right|^2\!\!
\vspace{0.15cm}\\ \displaystyle \hspace{1.25cm}
 =\frac{1}{2}\left\{\,1+\cos2\ellipin
 \cos2[\,\azaxeli+\eulerc(\vkond)-\Eularop\,]\,\right\},
\end{array}\hspace{0.75cm}
\label{eq:pislpb} \end{eqnarray}
 whatever $\helicin$, where $\eulerc(\vkond)$ is the rotation angle
 of the ellipse axes due to diffraction.
 From (\!\!~\ref{eq:pislpb}), we have:
\begin{eqnarray} \eulerc(\vkond)=\Eularop\!-\!\azaxeli
 +\frac{1}{2}\arccos\frac{\,2\left|\,\ilkeuz\debpscal\!\xpol
 \midpscal\etaelli^{(\spinun)}_{\textup{out}}(\vkond)\!\finpscal
 \right|^2\!\!-1\,}{\cos 2\ellipin}, \hspace{0.75cm}
\label{eq:pislpc} \end{eqnarray}
 where $\vkond=\vkond(\mkond,\eulanuy,\eulapre)$.
 Therefore, the measurement of the probability $\left|\,\ilkeuz\debpscal\xpol
 \midpscal\etaelli^{(\spinun)}_{\textup{out}}(\vkond)\finpscal\right|^2$
 as a function of $\mkond,\eulanuy,\eulapre$ makes it possible to fit the
 function $\eulerc[\vkond(\mkond,\eulanuy,\eulapre)]$ to the experimental data
 (provided that $\ellipin\neq\pi/4$).
 From (\!\!~\ref{eq:rotkok}) and (\!\!~\ref{eq:fapendi}),
 its expected value is zero for $\eulanuy=\eulapre=0$.
\\

 In the case of a linear polarization ($\ellipin=0$),
 the final polarization is also linear in the direction defined by the angle
 $\azaxeli+\eulerc(\vkond)$ (Eq. (\!\!~\ref{eq:inelpog})).
 Assuming that the maximum transmission axis of the analyzer is the axis
 $\oprotos(0,0,\Eularop)\,\cx[\vkond]$,
 the device can be rotated around $\cz[\vkond]$ so as to find the angle
 $\Eularop_{1}(\vkond)$ such that $\left|\,\,^{}_{\!\!\vkond,\Eularop_{1}(\vkond)
 \!\!\!}\debpscal\!\xpol\midpscal\etaelli^{(\spinun)}_{\textup{out}}(\vkond)
 \!\finpscal\right|^2=1$. Then, (\!\!~\ref{eq:pislpc}) leads to:
 $\eulerc(\vkond)=\Eularop_{1}(\vkond)-\azaxeli$.

}

}

\section{Conclusion}
{
\label{sec:conclu}

 It is possible to construct a model based exclusively on quantum mechanics
 to describe the Fraunhofer diffraction by a diaphragm.
 In the model presented here, the quantum concept of measurement was used,
 within the framework of the S-matrix formalism, to describe the passage
 of the particles through the aperture.
 The notion of projector had to be generalized by that
 of "filtering operator" in order to obtain a description of the measurement
 compatible with the Huygens-Fresnel principle.
 Then, because of kinematics,
 it was necessary to assume that the passage of the particle through
 the aperture is described by a "double measurement" starting with
 the measurement of position (which creates a localized transitional state
 of indeterminate energy) and ending with an energy-momentum measurement
 (which creates the final state with the same energy as the initial state).
 
 The model suggests that the wavelets involved in the Huygens-Fresnel principle
 are emitted from several neighboring wavefronts distributed along
 the longitudinal direction in the aperture region.
 These wavefronts contribute with different weights to the
 amplitude of the diffracted wave and the width of their distribution,
 not known a priori, can be fitted to the data from measurement of the intensity
 as a function of the diffraction angle.
 If this width is large enough,
 a significant damping of the intensity at large angles is predicted.
 A direct experimental study of the Huygens-Fresnel principle
 is therefore possible.
 Moreover, the model provides predictions concerning the still little explored
 region of large diffraction angles.
 In particular, it predicts the decrease in intensity towards zero
 at 90$^{\circ}$, contrary to most of the scalar theories of wave optics.
 Finally, in the case of light in single-photon states and for an incident
 monochromatic plane wave, the model predicts that the transfer of momentum
 between the photon and the diaphragm conserves the probabilities of the
 circular polarizations but can cause a phase shift between the amplitudes
 of the associated helicity states.
 For an initial state elliptically polarized,
 the conservation of the ellipticity and of the handedness is predicted.
 The phase shift between the amplitudes of the helicity states
 corresponds to a rotation of the axes of the ellipse.
 The angle of this rotation depends on the diffraction angles
 and is not known a priori.
 Its values can be fitted to the data from measurements of the polarization
 of the photons detected beyond the diaphragm.
 It would thus be possible to get information
 on how diffraction modifies the polarization of light.
\\

}

\section*{Acknowledgements}
{
 I would like to thank M. Besan\c{c}on, F. Charra and A. Rosowsky
 for their helpful suggestions and comments on the manuscript.
 I am also grateful to the late P. Roussel for his
 advice at the start of this work.
 I am thankful to the team of the IRAMIS-CEA/SPEC/LEPO:
 F. Charra, L. Douillard, C. Fiorini and S. Vassant,
 for fruitful discussions and achievement of preliminary tests
 for measurement of diffraction at large angles.
 This work was supported by the IRFU-CEA/DPhP.

}

{
\bibliographystyle{}

}

\end{document}